\documentclass[
]{jss}

\usepackage[utf8]{inputenc}
\usepackage{xr} 
\usepackage{hyperref} 

\makeatletter
\newcommand*{\addFileDependency}[1]{
  \typeout{(#1)}
  \@addtofilelist{#1}
  \IfFileExists{#1}{}{\typeout{No file #1.}}
}
\makeatother

\newcommand*{\myexternaldocument}[1]{%
    \externaldocument{#1}%
    \addFileDependency{#1.tex}%
    \addFileDependency{#1.aux}%
}

\usepackage{xr} 
\myexternaldocument{supplementary_material}

\providecommand{\tightlist}{%
  \setlength{\itemsep}{0pt}\setlength{\parskip}{0pt}}

\author{
Christian Capezza\\Univ. of Naples Federico II\\Corresponding author:\\ christian.capezza@unina.it \And
Fabio Centofanti\\Univ. of Naples Federico II \And
Antonio Lepore\\Univ. of Naples Federico II \AND
Alessandra Menafoglio\\Politecnico di Milano \And
Biagio Palumbo\\Univ. of Naples Federico II \And
Simone Vantini\\Politecnico di Milano
}
\title{funcharts: Control charts for multivariate functional data in
\proglang{R}

\large{\textbf{This is an original manuscript of an article published by Taylor \&
Francis in \textit{Journal of Quality Technology} on 19 July 2023, available at: https://doi.org/10.1080/00224065.2023.2219012}}
}

\Plainauthor{Christian Capezza, Fabio Centofanti, Antonio Lepore, Biagio Palumbo, Alessandra Menafoglio, Simone Vantini}
\Plaintitle{funcharts: Control charts for multivariate functional data
in R}
\Shorttitle{funcharts: Control charts for multivariate functional data
in R}

\Abstract{
Modern statistical process monitoring (SPM) applications focus on profile monitoring, i.e., the monitoring of process quality characteristics that can be modeled as profiles, also known as functional data.
Despite the large interest in the profile monitoring literature, there is still a lack of software to facilitate its practical application.
This article introduces the \pkg{funcharts} \proglang{R} package that implements recent developments on the SPM of multivariate functional quality characteristics, possibly adjusted by the influence of additional variables, referred to as covariates.
The package also implements the real-time version of all control charting procedures to monitor profiles partially observed up to an intermediate domain point.
The package is illustrated both through its built-in data generator and a real-case study on the SPM of Ro-Pax ship CO\textsubscript{2} emissions during navigation, which is based on the \texttt{ShipNavigation} data provided in the Supplementary Material. 
}

\Plainkeywords{functional data analysis, statistical process
control, profile monitoring, multivariate functional linear
regression, R}

\Address{
    }

\usepackage{caption}

\usepackage{amsmath,bm,amsfonts,bbm}

\begin{document}

\newpage

\section{Introduction}
\label{sec_intro}

Recent acquisition technologies are based on multiple and high-frequency sensors installed on board of devices or machines and allow the acquisition of a large amount of data about many process variables and quality characteristics. 
In this context, statistical process monitoring (SPM)  calls for methods able to model and analyze quality characteristics in the form of multiple curves, i.e., continuous functions defined on compact sets often referred to as functional data or profiles. 
For details on functional data analysis (FDA), the reader is referred to \cite{ramsay2005functional, horvath2012inference, ferraty2006nonparametric, kokoszka2017introduction}.
The large availability of functional data is leading to a growing interest in profile monitoring \citep{noorossana2011statistical}, which aims to check the stability over time of the process based on observations of one or more functional quality characteristics of interest, that is, as in the traditional SPM, to test whether or not the process is affected also by special and not only by common causes of variations. 
In the former case, the process is said to be out of control (OC), while in the latter case, it is said to be in control (IC). 
One of the first reviews on this topic is the paper of \cite{woodall2004using}.
For other relevant contributions, see \cite{jin1999feature}, \cite{colosimo2010comparison}, \cite{grasso2016using}, \cite{grasso2017phase}, \cite{menafoglio2018profile}, \cite{maleki2018overview}, and \cite{jones2021practitioners}.

The available approaches to profile monitoring are generally based on building control charts capable of detecting OC conditions, without necessarily considering the relationship between the quality characteristic and variables that may have an influence on it, referred to as covariates.
A few exceptions are the recent works of \cite{capezza2020control} and \cite{centofanti2020functional}, who propose methods to extend the regression control chart \citep{mandel1969regression} to functional data settings, where the quality characteristic is influenced by scalar or functional covariates, with the aim to improve the effectiveness of the monitoring strategy. 
In particular, they propose a control charting scheme based on the residuals of a linear regression with multivariate functional covariates used as regressors and the scalar \citep{capezza2020control} or functional quality characteristics used as the response \citep{centofanti2020functional}.

Surprisingly, despite the large interest in the profile monitoring literature, there is a lack of software specifically designed for this type of application in modern data-rich scenarios. 
The \pkg{qcc} R package \citep{scrucca2017qcc}, which is one of the most popular for the SPM, focuses indeed on control charts for univariate and multivariate data, only.
Also in the context of batch process monitoring, the \pkg{dvqcc} \citep{valk2020dvqcc} R package provides a nice set of control charts based on the vector autoregressive model \citep{marcondes2020dynamic} related to monitoring multiple variables observed over time but requires aligned data, while often profiles are observed over irregular grids with a different number of discrete points for each curve.
Moreover, none of the aforementioned packages implement SPM methods able to account for one or more covariates, possibly functional, which may influence the quality characteristic of interest.

The \pkg{funcharts} \proglang{R} package introduced in this work and available on CRAN \citep{capezza2021funcharts} is instead the first off-the-shelf toolkit able to perform SPM  of functional quality characteristics alone or adjusted by the influence of covariates.
In particular, in the former case, we provide the implementation of the methodology proposed in \cite{colosimo2010comparison} for profile monitoring of univariate functional data based on functional principal component analysis and its extension to the case of multivariate functional data. 
Whereas, in the latter case, we implement the aforementioned methods proposed by \cite{capezza2020control} and \cite{centofanti2020functional}. 
The \pkg{funcharts} package focuses on the prospective monitoring of the process, referred to as Phase II. 
That is, given a clean data set with \(n\) observations assumed to be representative of IC process performance, referred to as reference data set, the objective is to build appropriate control charts able to issue an alarm when the process is OC based on future individual observations drawn from it, hereinafter referred to as OC observations. 
Accordingly, we will refer to any future individual observations drawn from the IC process as IC observations.
 
In all the considered settings, the \pkg{funcharts} package implements also a real-time version of the functional control charts by accepting as input also profiles that are partially observed up to an intermediate domain point and adapting the real-time monitoring originally proposed by \cite{capezza2020control} in the case of scalar response to the case of functional response.


The remainder of this article is organized as follows. Section \ref{sec_overview} briefly overviews the methods that are implemented in the \pkg{funcharts} package. Comprehensive usage of the \pkg{funcharts} package is covered with a built-in simulation example in Section \ref{sec_using} and a real-case study in the monitoring of CO$_{2}$ emissions of a Ro-Pax ship during navigation in Section \ref{sec_realcasestudy}. Section \ref{sec_conclusions} concludes this article.

\section{Overview of the methods implemented in funcharts}
\label{sec_overview}

This section provides details of the methods implemented in \pkg{funcharts}. 
Section \ref{sec_mfpca} describes how multivariate functional principal component analysis (MFPCA) is performed to represent functional data.  
Section \ref{sec_methods_colosimo} reviews the methodologies for SPM of multivariate functional data without covariates and extends the method proposed by \cite{colosimo2010comparison} for univariate profiles to the multivariate setting. 
The control charts proposed by \cite{capezza2020control} and \cite{centofanti2020functional} are introduced in Section \ref{sec_methods_capezza} and \ref{sec_methods_centofanti}, respectively.
Finally, in Section \ref{sec_methods_realtime}, the real-time implementation of these methods is illustrated. 

\subsection{Multivariate functional principal component analysis}
\label{sec_mfpca}

Let \( \bm X_i = \lbrace(X_{i1}, \dots, X_{iP})^\top\rbrace_{i=1,\dots,n}\) denote \(n\) observations of a vector of \(P\) functional variables,  in the Hilbert space \(L^2 (\mathcal T)^P\), whose components $X_{i1}, \dots X_{iP}$ belong to $L^2(\mathcal T)$, the space of square integrable functions defined on the compact interval $\mathcal T$.
More specifically, the \pkg{funcharts} package assumes that these functional data are densely observed on a set of discrete grid points (usually in time or space) and can be represented through cubic B-spline basis expansion with order four and equally spaced knots. 
The B-spline basis system has shown to enjoy remarkable theoretical as well as computational properties in the problem of recovering a smooth signal from noisy measurements \citep{de1978practical,wahba1990spline,green1994nonparametric}. 
To get a functional observation from the discrete grid points, a large number of basis functions is chosen and basis coefficients are estimated using penalized least squares. 
The penalty term is the integrated squared second derivative with smoothing parameter selected through generalized cross-validation (GCV) \citep{ramsay2005functional}.

MFPCA \citep{ramsay2005functional,jacques2014model,happ2018multivariate} is then used to get a finite representation of functional data.
Moreover, to take into account the possible different ranges of variation of the $P$ functional variables, these are empirically standardized, as in \cite{chiou2014multivariate}, i.e., by firstly subtracting pointwise from each original variable the corresponding sample mean function and then dividing the result pointwise by the relative sample standard deviation function.
Then, MFPCA consists in finding the eigenvalues \(\lambda_m\) and the corresponding eigenfunctions \(\bm \psi_m = (\psi_{m1}, \dots, \psi_{mP})^\top \in L^2 (\mathcal T)^P\), with \(1 \leq m \leq n-1\), as the solutions of the eigenvalue equation 
\begin{equation}
\int_{\mathcal T} \bm V(s,t) \bm \psi_{m} (s) ds = \lambda_m \bm \psi_{m} (t), \quad \forall t \in \mathcal T,
\end{equation} 
where $\bm V(s,t)=\frac{1}{n-1} \sum_{i=1}^n \bm X_{i}(s) \bm X_{i}(t)^\top$ is the empirical covariance function of the multivariate functional data.
Eigenvalues are arranged in a non-increasing order, i.e.~\(\lambda_1 \geq \lambda_2 \geq \dots \geq 0\). 
Eigenfunctions are orthogonal to each other, i.e., \(\sum_{p=1}^P \int_{\mathcal T} \psi_{m_1 p} (t) \psi_{m_2 p} (t) dt = \mathbbm{1}(m_1=m_2)\), where \(\mathbbm{1}(\cdot)\) is the indicator function, and are also known as multivariate functional principal components, or principal components (PCs). 
For computational details, the reader is referred to Section 8.4.2 of \cite{ramsay2005functional}. 
By projecting each observation onto the the PCs, we get the multivariate functional PC scores, or simply scores \(\xi_{im}=\sum_{p=1}^P \int_{\mathcal T} X_{ip}(t) \psi_{mp} (t) dt\), which accordingly satisfy \(\sum_{i=1}^n \xi_{im}=0\) and \(\sum_{i=1}^n \xi_{im_1} \xi_{im_2} / (n-1) = \lambda_{m_1} \mathbbm{1}(m_1=m_2)\).

A finite representation of the original data is obtained by retaining a finite subset of the PCs. 
In the following, we consider the most popular choice to retain the first \(M\) PCs, i.e., the ones associated with the largest \(M\) eigenvalues. 
Alternative choices are discussed in
\cite{capezza2020control}.
Thus, \(\bm X_i\) is approximated with \( \hat {\bm X_i} = (\hat X_{i1}, \dots, \hat X_{iP})^\top\), defined as
\begin{equation}
\label{eq_sof_pca}
\hat X_{ip} \left(t\right)=\sum_{m=1}^{M} \xi_{im}\psi_{mp}\left(t\right), \quad t \in \mathcal T, p=1,\dots,P.
\end{equation}

\subsection{Functional control charts for multivariate quality characteristics}
\label{sec_methods_colosimo}

Let us consider a vector of $P$ functional variables as a functional quality characteristics of interest to be monitored.
To this aim, for each observation \(\lbrace \bm X_i \in L^2 (\mathcal T)^P \rbrace_{i=1,\dots,n}\), two control charts are respectively built on the Hotelling's \(T^2\) and squared prediction error ($SPE$) statistics, defined as
\begin{equation}
\label{eq_t2_spe}
T_i^2=\sum_{m=1}^M \xi_{im}^2 / \lambda_m, \quad SPE_i = \sum_{p=1}^P \int_{\mathcal T} (X_{ip}(t) - \hat X_{ip}(t))^2 dt.
\end{equation}
where the $SPE$ statistic can be also calculated as $SPE_i = \sum_{M+1}^{n-1} \xi_m^2$. The Hotelling's \(T^2\) statistic monitors \(\hat {\bm X_i}\) defined in Equation \eqref{eq_sof_pca} through the retained PCs, while the $SPE$ statistic monitors the approximation error due to the dimension reduction.
While in \cite{colosimo2010comparison} control chart limits \(T^2_{lim}\) and \(SPE_{lim}\) are calculated based on the assumption that scores are normally distributed, in the \pkg{funcharts} package they are calculated non-parametrically by means of empirical quantiles of the reference data set distribution of the Hotelling's $T^2$ and $SPE$ statistics.
The \pkg{funcharts} package uses the Bonferroni correction \citep{hsu1996multiple} to constrain the family-wise error rate to be smaller than a desired value $\alpha \in (0,1)$ by setting the Type I error rate in each of the two control charts equal to \(\alpha/2\), that for each control chart is the nominal probability that an in-control point falls outside of the control limits.
Moreover, as shown in \cite{capezza2020control}, the Hotelling's \(T^2\) and \(SPE\) statistics can be decomposed as sums over the \(P\) functional variables. 
In this way, the contributions of the $p$-th variable, $p=1,\dots,P$, of the $i$-th observation, $i=1,\dots,n$, to each monitoring statistic can be, respectively, measured by $CONT_{T^2_i,p}$ and $CONT_{SPE_i,p}$ as follows
\begin{equation}
\label{eq_cont}
CONT_{T^2_i,p} = \sum_{m=1}^M \frac{\xi_{im}}{\lambda_m} \int_{\mathcal T} X_{ip}(t) \psi_{mp}(t) dt, \quad CONT_{SPE_i,p} = \int_{\mathcal T} (X_{ip} (t) - \hat X_{ip} (t))^2 dt.
\end{equation}
These quantities allow to identify which functional variables have been responsible of especially large values of a monitoring statistic and are then an important tool for fault detection.
\cite{capezza2020control} noted that  these contribution terms may have very different variability, for each $p=1,\dots,P$.
Therefore, large variable contributions are identified as those that exceed the relative empirical upper limits.

Let us denote by \(\bm X^* = (X^*_1, \dots, X^*_P)^\top\) a future  observation  to be monitored in Phase II.
After standardization with respect to the sample mean and standard deviation functions calculated on the reference data set, scores are calculated by projecting $\bm X^*$ onto the the PCs.
This allows one to calculate the approximation $\hat{\bm X}^*$ using Equation \eqref{eq_sof_pca} and the two monitoring statistics, which we denote as $T^{2*}$ and $SPE^*$, using Equation \eqref{eq_t2_spe}.
If either the $T^{2*}$ or $SPE^*$ monitoring statistic observation is larger than the corresponding control limit, an alarm is issued. 
Then, contributions to the OC monitoring statistic are calculated through Equation \eqref{eq_cont} and variables with a contribution that is larger than the corresponding limit need to be investigated.

\subsection{Functional control charts for a univariate quality characteristic adjusted by the influence of covariates}
\label{sec_supervised_method}

Differently from Section \ref{sec_methods_colosimo}, let us consider a univariate quality characteristic, in the form of a scalar or a function, to be monitored when functional covariates are available. 
After the quality characteristic is adjusted by the functional covariates, the control chart is improved because the monitoring is conditional on the observed value of the functional covariates.
Section \ref{sec_methods_capezza} and \ref{sec_methods_centofanti} show the control charts proposed by \cite{capezza2020control} and \cite{centofanti2020functional} for a scalar and functional quality characteristic, respectively.

\subsubsection{Control charts based on scalar-on-function regression}
\label{sec_methods_capezza}

\cite{capezza2020control} build control charts based on the scalar-on-function linear regression model, between observations of a scalar response variable, denoted by $y_i$, and observations of a vector of functional covariates \(\bm X_i = (X_{i1}, \dots, X_{iP})^\top \in L^2 (\mathcal T)^P\). 
That is, the model is defined as
\begin{equation}
\label{eq_sof_model}
y_i = 
\beta_0 + \sum_{p = 1} ^ P \int_\mathcal{T} {X}_{ip} (t) {\beta}_{p} (t) dt + \varepsilon_i, \quad i=1,\dots,n,
\end{equation} where \(\bm \beta = (\beta_1, \dots, \beta_P)^\top \in L^2 (\mathcal T)^P\) are the functional coefficients, \(\beta_0\) is the scalar intercept, and \(\varepsilon_1, \dots, \varepsilon_n\) are the independent error terms with identical distribution \(\mathcal N(0, \sigma^2)\). 
Without loss of generality, we assume that the functional covariates have already been standardized as shown in Section \ref{sec_mfpca}.
Because of the infinite dimensionality of the functional data, this model cannot be estimated directly using least squares. 
By applying MFPCA to the functional covariates for dimension reduction as shown in Section \ref{sec_mfpca}, functional covariates can be approximated by $\hat{\bm X}_i$ as in Equation \eqref{eq_sof_pca}. 
Moreover, eigenfunctions can be used for basis expansion of the functional coefficients, leading to the truncated version $\bm \beta^M = (\beta^M_1,\dots,\beta^M_P)^\top \in L^2 (\mathcal T)^P$
\begin{equation}
\label{eq_sof_bm}
\beta^M_p(t) = \sum_{m=1}^M b_{m} \psi_{mp} (t), \quad t \in \mathcal T, p = 1, \dots, P.
\end{equation}
In this way, model \eqref{eq_sof_model} is replaced by the following approximate model
\begin{equation}
\label{eq_sof_simplified}
y_i = \beta_0 + \sum_{m=1}^M \xi_{im} b_m + \varepsilon_i^M,
\end{equation}
where least-squares estimates are obtained as
\begin{equation}
\label{eq_sof_ls}
\hat\beta_0 = \frac{1}{n} \sum_{i=1}^n y_i, \quad \hat b_m = \sum_{i=1}^n y_i \xi_{im}/\sum_{i = 1}^n \xi_{im}^2, \quad m=1, \dots, M.
\end{equation}
Finally, the estimate $\hat {\bm \beta} = (\hat \beta_1, \dots, \hat \beta_P)^\top$ of $\bm \beta^M$ and then of $\bm \beta$ is obtained upon replacing $b_1 \dots, b_M$ in Equation \eqref{eq_sof_bm} with the least-squares estimates in Equation \eqref{eq_sof_ls}, and hence the prediction of the scalar response variable is given by 
\begin{equation}
\label{eq_sof_pred}
\hat y_i = \hat \beta_0 + \sum_{p=1}^P \int_{\mathcal T} X_{ip} (t) \hat \beta_p (t) dt.
\end{equation}
Given the estimated model, a future  observation of the scalar response variable \(y^*\) and the corresponding functional covariate vector \(\bm X^* = (X^*_1, \dots, X^*_P)^\top\) is monitored in Phase II through three control charts built on (\(i\)) the Hotelling's \(T^{2*}\) and (\(ii\)) \(SPE^*\) monitoring statistics, as shown in Section \ref{sec_methods_colosimo}, as well as (\(iii\)) the response prediction error \(y^* - \hat y^*\), with $\hat y^*$  obtained through Equation \eqref{eq_sof_pred}. The response prediction error control chart intends to monitor the scalar response variable conditionally on the functional covariates.
Given an overall Type I error probability $\alpha$, the Bonferroni correction is used by setting in each control chart the Type I error rate equal to \(\alpha/3\). 
The limits of the Hotelling's \(T^2\) and \(SPE\) control charts are calculated non-parametrically as described in Section \ref{sec_methods_colosimo}. 
Because of the normality assumption on the error terms, control limits of the control chart on the response prediction error are calculated as
\begin{equation}
\pm t_{n-M-1,1-\alpha/6} \hat \sigma \left( 1 + \frac{T^{2*}}{n-1} \right)^{1/2},
\end{equation}
where \(t_{\nu,\gamma}\) is the \(\gamma\)-quantile of the Student's distribution with \(\nu\) degrees of freedom and \(\hat \sigma^2 = \sum_{i=1}^n (y_i - \hat y_i)^2 / (n-M-1)\). 
If at least one of the three monitoring statistics exceeds the relative control limits, then an OC alarm is issued. The contributions to the Hotelling's \(T^2\) and \(SPE\) statistics presented in Section \ref{sec_methods_colosimo} can support the inspection of functional covariates  responsible of the OC alarm. In contrast, when an OC is issued by the response prediction error control chart,  causes should be possibly investigated outside of the set of variables included in the model as covariates.

\subsubsection{Control charts based on function-on-function regression}
\label{sec_methods_centofanti}
The response prediction error control chart presented in Section \ref{sec_methods_capezza} can be regarded as a particular case of the functional regression control chart (FRCC), which is a general framework for profile monitoring proposed by \cite{centofanti2020functional}. 
The FRCC is based on the following three main steps, in which one should define ($i$) the functional regression model to be fitted; ($ii$) the estimation method of the chosen model;  ($iii$)  the monitoring strategy of the functional residuals.
Specifically, \cite{centofanti2020functional} describe a particular implementation of the FRCC framework obtained by considering (step ($i$)) the  function-on-function linear regression
model \begin{equation}
\label{eq_fof}
y_i(t) = \beta_0 (t) + \sum_{p = 1} ^ P \int_\mathcal{S} {X}_{ip} (s) {\beta}_{p} (s, t) ds + \varepsilon_i (t), \quad i=1,\dots,n, \quad t \in \mathcal T,
\end{equation} 
where \(y_i \in L^2(\mathcal T)\), differently from \cite{capezza2020control}, is allowed to be a functional response variable. 
The multivariate functional covariates \(\bm X_i = (X_{i1}, \dots, X_{iP})^\top \in L^2 (\mathcal S)^P\), i.e., are allowed to be defined on a compact domain $\mathcal S$, not necessarily equal to $\mathcal T$, and \(\bm \beta = (\beta_1, \dots, \beta_P)^\top \in L^2 (\mathcal S \times \mathcal T)^P\) are the functional coefficients, \(\beta_0 \in L^2 (\mathcal T)\) is the functional intercept. 
In Equation \eqref{eq_fof}, \(\varepsilon_1, \dots, \varepsilon_n\) denote the independent functional error terms with zero mean and covariance function $V_\varepsilon\ \in L^2(\mathcal S \times \mathcal T)$. 
Without loss of generality, we assume that both the functional response and covariates have already been empirically standardized as shown in Section \ref{sec_mfpca}.
Accordingly,  the functional intercept is set equal to zero in model \eqref{eq_fof}.
Then,  the step ($ii$) is performed by applying MFPCA to both the set of functional covariates and to the response function separately, as described in Section \ref{sec_mfpca}, yielding respectively eigenfunctions \(\bm \psi^X_{l} = (\psi^X_{l1}, \dots \psi^X_{lP})^\top \in L^2 (\mathcal S)^P\) and $ \psi^Y_m \in L^2(\mathcal T)$, to which correspond eigenvalues \(\lambda^X_l\) and \(\lambda^Y_m\). 
The covariates $X_{ip}$ and the response $Y_i$, $i=1,\dots,n$, can be then  approximated by the relative truncated version
\begin{equation}
\hat X^L_{ip}(s) = \sum_{l=1}^L \xi^X_{il} \psi^X_{lp}(s), \quad p=1,\dots, P,  s \in \mathcal S, \quad \hat Y^M_{i}(t) = \sum_{m=1}^M\xi^Y_{im} \psi^Y_{m}(t), \quad t \in \mathcal T,
\end{equation}
where $\xi^Y_{im}=\int_{\mathcal T} Y_{i}(t) \psi_{m}^Y (t) dt$, $\xi^X_{il}=\sum_{p=1}^P \int_{\mathcal S} X_{ip}(s) \psi_{lp}^X (s) ds$, and $L$ and \(M\) denote the number of PCs retained to represent $X_{ip}$ and  $Y_i$, respectively.
Moreover, in the same fashion,  the functional coefficients $\bm \beta$ in Equation \eqref{eq_fof},  can be represented by the truncated version $\bm \beta^{LM} = (\beta^{LM}_1, \dots, \beta^{LM}_P)^\top$, where each component is obtained as
\begin{equation}
\label{eq_fof_lm}
\beta_p(s, t)^{LM} = \sum_{l=1}^L \sum_{m=1}^M b_{lm} \psi^X_{mp} (s) \psi^Y_l(t), \quad s \in \mathcal S, t \in \mathcal T, p = 1, \dots, P.
\end{equation}
Then, model \eqref{eq_fof} is replaced by the following approximate model
\begin{equation}
\label{eq_fof_simplified}
\xi^Y_{im} = \sum_{l=1}^L \xi^X_{il} b_{lm} + \epsilon_{im}, \quad m=1,\dots,M.
\end{equation}
Basis coefficients $b_{lm}$ are estimated through least squares as \(\hat b_{lm} = \sum_{i = 1}^n (\xi^Y_{im} \xi^X_{il}) / \sum_{i = 1}^n (\xi^X_{il})^2\).
The latter can be plugged into Equation \eqref{eq_fof_lm} in place of $b_{lm}$ to obtain $\hat{\beta}_p(s,t)^{LM}$, and thus  the estimate $\hat {\bm \beta} = (\hat \beta_1, \dots, \hat \beta_P)$ of $\bm \beta$, where $\hat{\beta}_p=\hat{\beta}_p(s,t)^{LM}$, $s\in\mathcal{S}$, $t\in\mathcal{T}$, $p = 1, \dots, P$.
Predictions of the functional response $Y_i(t)$ are obtained as 
\begin{equation}
\hat Y_i(t) = \sum_{p = 1} ^ P \int_\mathcal{S} {X}_{ip} (s) {\hat \beta}_{p} (s, t) ds, \quad t \in \mathcal T.
\end{equation} 
Finally, in the last step ($iii$) of the FRCC, the most standard choice proposed by \cite{centofanti2020functional} is to represent the functional residuals 
\begin{equation}
\label{eq_fof_res}
e_i (t) = Y_i (t) - \hat Y_i(t), \quad t \in \mathcal T,
\end{equation} 
with respect to the  first $K$ retained functional principal components.
$K$ can be chosen accordingly to a given percentage, say  95\%  or  99\%, of total variability explained.
Eigenfunctions, eigenvalues, and scores are respectively denoted by \(\psi^e_k \in L^2(\mathcal T)\),  \(\lambda^e_k\), and  \(\xi^e_k\), with \(k=1,\dots,K\).
An alternative choice in step ($iii$)  is based on a studentized version of the functional residuals  suggested by \cite{centofanti2020functional}, when $N$ is not particularly large and in presence of covariate mean shifts. 
The studentized functional residuals are defined as 
\begin{equation}
\label{eq_stud_res}
e_{i,\text{stu}} (t) = \frac{Y_i (t) - \hat{Y}_i(t)}{\left(\hat{v}_{\varepsilon} (t) + \bm \psi^Y(t)^\top \hat {\bm \Sigma}_{\varepsilon} \bm \psi^Y(t) \sum_{l=1}^L (\xi^X_{il})^2 / \lambda^X_{l} \right)^{1/2}}, \quad t \in \mathcal T,
\end{equation}
where  \(\bm \psi^Y = (\psi_{1}^Y, \dots, \psi_{M}^Y)^\top\), \(\bm \epsilon_i = (\epsilon_{i1}, \dots, \epsilon_{iM})^\top\), \(\hat v^2_{\varepsilon}(t) = \sum_{i=1}^n e_i(t)^2/ (n-1)\), $t \in \mathcal T$ is the estimator of the variance of the residual function,  and \(\hat {\bm \Sigma}_{\varepsilon} = \sum_{i=1}^n \bm \epsilon_i \bm \epsilon_i^\top / n\) is the estimator of the covariance matrix of the errors \(\epsilon_{im}\) in Equation \eqref{eq_fof_simplified}.
To conclude step ($iii$), apart from the residual version used, the monitoring strategy of the residuals is based on Hotelling's \(T^2\) and \(SPE\) control charts constructed on the $K$ scores  \(\xi^e_k\).
Then, given the estimated model, a future observation of the functional response variable and the corresponding functional covariates is monitored in  Phase II  by calculating the functional residual (as in Equation \eqref{eq_fof_res} or  Equation \eqref{eq_stud_res}); by projecting it onto  $\psi^e_k$ to obtain future scores; and calculating the Hotelling's \(T^2\) and \(SPE\) monitoring statistics. 
If at least one of the two monitoring statistics is out of the control limits, then an OC alarm is issued. 

\subsection{Real-time functional control charts}
\label{sec_methods_realtime}

The methods presented so far can be extended to monitor in real time functional data that are partially observed up to an intermediate, e.g., the current, domain point.
Let the functional domain be the interval $\mathcal T = (a, b)$ for all observations and denote with $t^{*} \in \mathcal T$ the intermediate domain point up to which a future functional observation is available.
The reference data set up to $t^{*}$ is obtained by truncating the reference observations of the functional data such that the functional domain is $(a, t^{*})$.
This is used to estimate the reference distribution of the monitoring statistics at $t^{*}$ and hence, the limits of each control chart, as described in Sections \ref{sec_mfpca}, \ref{sec_methods_colosimo} and \ref{sec_supervised_method}.
These limits are used to compare the monitoring statistics observation at $t^{*}$, which is calculated on the future functional observation. 
As before, an alarm is issued if at least one monitoring statistic plots outside of the corresponding control limits. 
This strategy has been proposed for the first time by \cite{capezza2020control}, which the reader is referred to for details, to perform real-time monitoring of functional data in the scalar-on-function regression case of Section \ref{sec_methods_capezza}.

\section{Using funcharts}
\label{sec_using}

In this section, the \pkg{funcharts} package is illustrated with reference to the methods described in Section \ref{sec_overview}. 
It uses the \pkg{fda} package \citep{ramsay2021fda} to get functional data from the discrete points and to perform MFPCA. 
In particular, Section \ref{how_to_simulate} illustrates how the \pkg{funcharts} package built-in  data generator works in simulating  functional data  drawn form either the IC process or from a given OC scenario in terms of process mean shift. 
Section \ref{mfd_class} shows the
\texttt{mfd} class to deal with multivariate functional data. 
Sections \ref{sec_using_colosimo} and \ref{sec_using_supervised}
apply the \pkg{funcharts} package to implement the methods described in Sections \ref{sec_methods_colosimo} and \ref{sec_supervised_method}, respectively. 
Finally, the real-time versions of the main functions of the \pkg{funcharts} package are given in Section \ref{sec_using_realtime} for the implementation of the real-time monitoring mentioned in Section \ref{sec_methods_realtime}.

\hypertarget{how_to_simulate}{%
\subsection{Simulating functional data}\label{how_to_simulate}}

The data generation mechanism is based on the simulation study shown in \cite{centofanti2020functional} for the function-on-function regression case and works similarly for the scalar-on-function regression case.
Trivially, to get a multivariate functional data set to be used, e.g., as in Section \ref{sec_methods_colosimo}, one can simply extract the generated observations of the functional covariates discarding the response. 

The function \texttt{sim\_funcharts()} generates an output list with three elements.
The first one, \texttt{datI}, contains the reference data set of $n_{I}$ observations to be used for model fitting.
The second, \texttt{datI\_tun}, contains additional $n_{{tun}}$ IC observations to be used to estimate control chart limits, given the model estimated using \texttt{datI}. 
The third, \texttt{datII}, contains a data set to be monitored in Phase II, which is made up of $n_{II}$ observations, splitted into three groups of equal size.
By default, $n_{I}=n_{{tun}}=1000$ and $n_{II}=60$.
The first group is made of data from an IC process, the second group is made up of OC observations with a modest mean shift, while the third group is made up of observations with more severe mean shifts.
Four types of shifts are available, labelled as A, B, C, D as in \cite{centofanti2020functional}. 
Shift A is representative of a change in the mean function curvature, whereas shifts B and C represent slope modification and translation of the profile pattern, respectively. 
Shift D consists of both curvature and slope modifications of the mean function.
The degree of severity is controlled by a parameter $d$.
For additional details, we refer to \cite{centofanti2020functional}.
Moreover, details on \texttt{simulate\_mfd()} and the data generation are available in Supplementary Material S1.

Each data set contains observations of three functional covariates \texttt{X1}, \texttt{X2} and \texttt{X3}, a functional response \texttt{Y} and a scalar response \texttt{y\_scalar}, where functional variables are observed on 150 discrete point equally spaced on the functional domain \(\mathcal{T} = (0, 1)\).
To change the simulation scenario, such as the sample size or the mean shift types, the \texttt{simulate\_mfd()} function, on which \texttt{sim\_funcharts()} relies, can be used.
An \proglang{R} code example to generate the specific data sets analyzed in the next subsections is as follows

\begin{CodeChunk}
\begin{CodeInput}
R> library(funcharts)
R> set.seed(0); d <- sim_funcharts()
\end{CodeInput}
\end{CodeChunk}

\hypertarget{mfd_class}{%
\subsection{The mfd class}\label{mfd_class}}

The \pkg{funcharts} package provides the \texttt{mfd} class for
multivariate functional data, which inherits  the \texttt{fd} class from the \pkg{fda} package and provides some additional features. 
Additional details on the \texttt{mfd} class are available in Supplementary Material S2.
Let us show how to get functional data with the generated data sets. 
Since \texttt{sim\_funcharts()} provides lists of matrices as output, we use the
\texttt{get\_mfd\_list()} function as follows

\begin{CodeChunk}
\begin{CodeInput}
R> mfdI <- get_mfd_list(d$datI[1:4]$)
R> mfdI_tun <- get_mfd_list(d$datI_tun[1:4]$)
R> mfdII <- get_mfd_list(d$datII[1:4]$)
\end{CodeInput}
\end{CodeChunk}

\begin{figure}

\centering \includegraphics[width=.666667\textwidth]{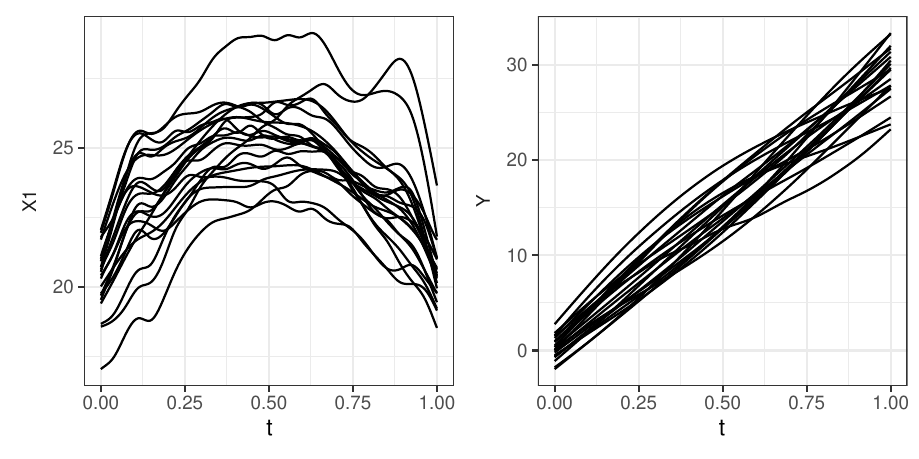}

\caption{\label{fig:plot_mfd}Plot of 20 simulated observations from the reference data set of the first functional covariates and the functional response.}
\end{figure}

For plotting objects of class \texttt{mfd}, the \pkg{funcharts} package provides the \texttt{plot\_mfd()} function based on the \pkg{ggplot2} package. 
Figure \ref{fig:plot_mfd} shows the plot of the first 20 replications of the functional variables \texttt{X1} and \texttt{Y} for a reference data set.

\begin{CodeChunk}
\begin{CodeInput}
R> plot_mfd(mfdI[1:20 , c("X1", "Y")])
\end{CodeInput}
\end{CodeChunk}

\hypertarget{sec_using_colosimo}{%
\subsection{Control charts for multivariate functional data}\label{sec_using_colosimo}}

In this section we show how to use the \pkg{funcharts} package to build control charts in the setting described in Section \ref{sec_methods_colosimo} with the simulated data. 
Note that in the second group of Phase II data, the third functional variable has a moderate mean shift of type A with severity \(d=20\), while in the third group of Phase II data the third functional variable has a more severe mean shift of type A with severity \(d=40\).

For class \texttt{mfd}, the \pkg{funcharts} package provides the
\texttt{pca\_mfd()} function to perform MFPCA, which is a wrapper to \texttt{pca.fd()}
from the \pkg{fda} package. Note that \texttt{pca\_mfd()} is called
internally by the functions \texttt{control\_charts\_pca()},
\texttt{sof\_pc()}, and \texttt{fof\_pc()}, that will be shown in the next sections,
before building the corresponding control charts. 
The \pkg{funcharts} provides also the
\texttt{plot\_pca\_mfd()} function to plot the eigenfunctions selected by where the argument
\texttt{harm}. For example, in Figure
\ref{fig:eigenfunctions} we plot the first two eigenfunctions of the covariance operator of the three functional covariates, calculated on the reference data set observations in the object \texttt{mfdI}.

\begin{CodeChunk}
\begin{CodeInput}
R> pca <- pca_mfd(mfdI[, c("X1", "X2", "X3")])
R> plot_pca_mfd(pca, harm = 1:2)
\end{CodeInput}
\begin{figure}

{\centering \includegraphics[width=\textwidth]{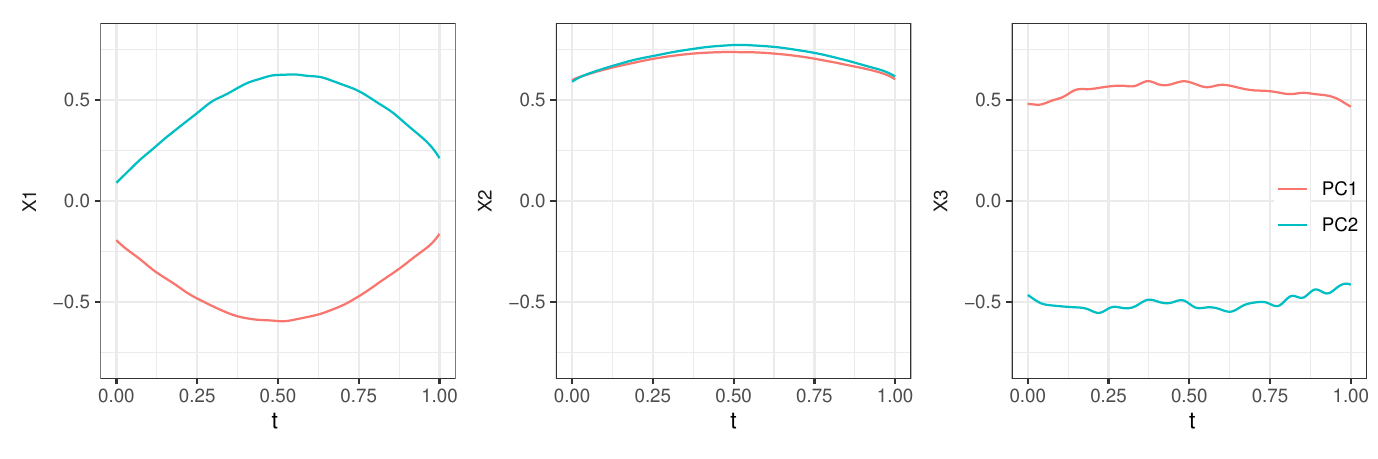} 

}

\caption{\label{fig:eigenfunctions}Plot of the first two eigenfunctions (PC1 and PC2) $\bm \psi_1(t) = (\psi_{11}(t), \psi_{12}(t), \psi_{13}(t))^\top$ and $\bm \psi_2(t) = (\psi_{21}(t), \psi_{22}(t), \psi_{23}(t))^\top$ of the empirical covariance operator of the three functional covariates X1, X2 and X3, reported in each of the three panels, respectively.}
\end{figure}
\end{CodeChunk}

At this stage, the \pkg{funcharts} package can be suitably used to get 
a data frame with all the information required to plot the  Phase II  Hotelling's \(T^2\) and  \(SPE\)  control charts as explained in Section \ref{sec_methods_colosimo}.

The main arguments of \texttt{control\_charts\_pca()} are:

\begin{itemize}
\tightlist
\item
  \texttt{pca},  the fitted MFPCA model obtained as output of
  \texttt{pca\_mfd()} on the reference data set.
\item
  \texttt{components}, a vector of the eigenfunctions to be retained in the MFPCA model.
\item
  \texttt{tuning\_data}, an optional data set of IC observations to be used for control charts limits estimation.
  If no argument is provided, limits will be calculated as quantiles of the statistics computed on the reference data set.
\item
  \texttt{newdata}, the \texttt{mfd} object containing the simulated  observations to be monitored in Phase II.
\item
  \texttt{alpha}, a numeric list containing   the Type I errors to be used to estimate  the   Hotelling's \(T^2\) and  \(SPE\) control chart limits.
  The default value is   \texttt{list(T2\ =\ 0.025,\ spe\ =\ 0.025)}, which uses the Bonferroni correction to ensure that the overall Type I error \(\alpha\) is not   larger than \(0.05\).
\end{itemize}

Then, the \texttt{plot\_control\_charts()} function can be used with the
 data frame obtained to plot the three control charts shown in Figure
\ref{fig:pca_cc}. In the following, we select the first components that
explain at least 95\% of the total variability in the data, as the
default option. We  add two vertical dashed lines to distinguish 
three groups of 20 Phase II data, generated under different scenarios. 
Note that \texttt{plot\_control\_charts()} relies on  the \pkg{patchwork} package
\citep{pedersen2020patchwork} to arrange the \pkg{ggplot} objects, and  to add a vertical line to all plots through
the operator \texttt{\&}, as follows

\begin{CodeChunk}
\begin{CodeInput}
R> cclist_pca <- control_charts_pca(
+   pca = pca,
+   tuning_data = mfdI_tun[, c("X1", "X2", "X3")],
+   newdata = mfdII_pca)
R> plot_control_charts(cclist_pca) &
+   geom_vline(aes(xintercept = c(20.5)), lty = 2) &
+   geom_vline(aes(xintercept = c(40.5)), lty = 2)
\end{CodeInput}
\begin{figure}

{\centering \includegraphics[width=\textwidth]{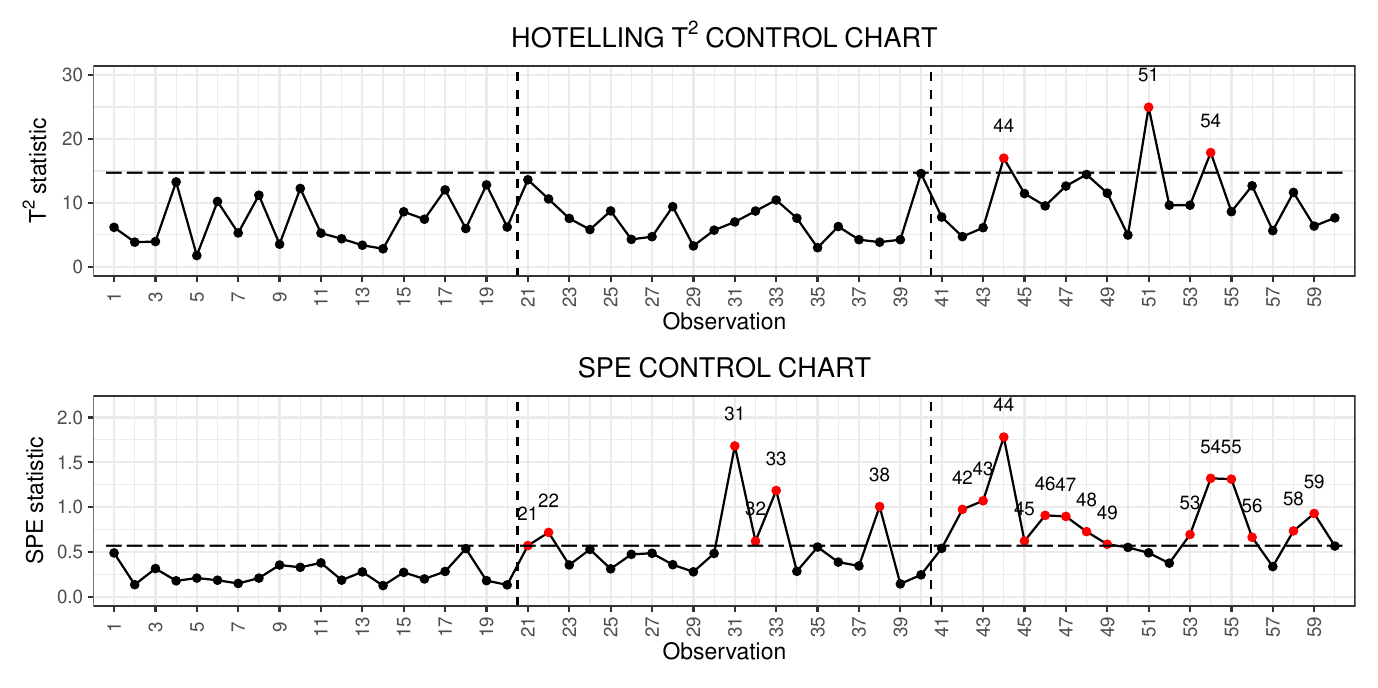} 

}

\caption{\label{fig:pca_cc}Control charts for multivariate functional data without covariates. Vertical lines separate the three groups of simulated Phase II data, generated as described in Section \ref{how_to_simulate}. Horizontal lines correspond to control chart limits, points correspond to the observations of the monitoring statistics for each Phase II observation. Red points highlights monitoring statistics that are out of the corresponding control limits; above each of them is reported the corresponding OC observation index.}
\end{figure}
\end{CodeChunk}

From Figure \ref{fig:pca_cc}, we see that the first group of  20 Phase II observations, which have been generated from the IC distribution rightly do not show up as OC points. Whereas, the second group of 20 Phase II observations, where  a mean shift on the third functional variable was simulated contains OC points.
Ultimately, the last group of 20 Phase II observations,
where a more severe shift on the third functional variable was set, shows up as OC, except for 3 points. The \pkg{funcharts} package allows one
to investigate also  which original variable contributes the most and so may be responsible of any OC signal, through the \texttt{cont\_plot()} function, which  displays signal  contributions to the Hotelling's \(T^2\) and \(SPE\) monitoring statistics. 
For example, let us consider observation 59, which is OC in the \(SPE\) control chart. 
The main arguments
of \texttt{cont\_plot()} are:

\begin{itemize}
\tightlist
\item
  \texttt{cclist},  the data frame returned by
  \texttt{control\_charts\_sof\_pc}.
\item
  \texttt{id\_num},  the index of the observation
  whose contributions are to be plotted.
\item
  \texttt{which\_plot}, a vector containing the desired
  contribution to be plotted, where \texttt{"T2"} denotes the
  Hotelling's \(T^2\) statistic and \texttt{"spe"} denotes the \(SPE\)
  statistic.
  \end{itemize}
  The contributions are shown as bar plots, where upper
  control limits are added as lines estimated on the Phase I tuning
  data. Contributions that exceed the limits are plotted in red color otherwise bar are plotted in gray.
  The
  contributions in Figure \ref{fig:cont_plot_sof} show  the second and third
  functional variables as the main responsible of the OC in
  the \(SPE\) statistic, thus correctly signaling the simulated mean shift.
  Note that, even though we simulated a mean shift in varaible $X_3$, contribution of the functional variable $X_2$ is also particularly large. This is reasonable because variable $X_3$ affects 
  the entire covariance structure in the data, which can turn into large
  contributions to the monitoring statistics also from other variables
  correlated with it.

\begin{CodeChunk}
\begin{CodeInput}
R> cont_plot(cclist = cclist_pca, id_num = 59, which_plot = "spe")
\end{CodeInput}
\begin{figure}

{\centering \includegraphics[width=0.37\textwidth]{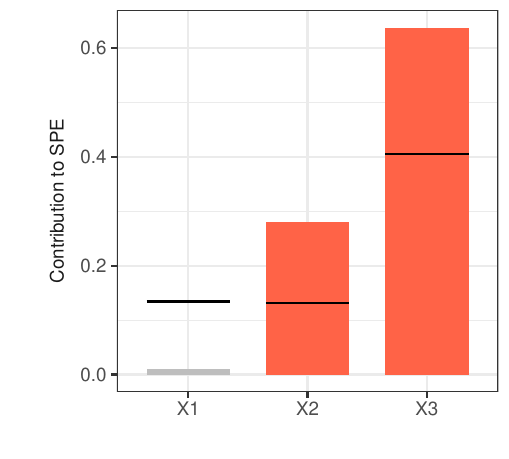} 

}

\caption{\label{fig:cont_plot_sof}Contribution of the functional covariates to the Hotelling's $SPE$ statistic for the Phase II observation 59.  Contributions exceeding upper control limits depicted by horizontal lines   are plotted in red.}
\end{figure}
\end{CodeChunk}

Finally, we can also plot any future observation against the reference
data set by using the \texttt{plot\_mon()} function, as shown in Figure
\ref{fig:plot_mon_pca}. Its main arguments are:

\begin{itemize}
\tightlist
\item
  \texttt{cclist}, the data frame returned by
  \texttt{control\_charts\_pca}.
\item
  \texttt{fd\_train}, an \texttt{mfd} object containing the data set
  to be plotted in the background in gray. Usually the reference data set
  of functional variables representing the IC performance is provided.
\item
  \texttt{fd\_test}, an \texttt{mfd} object containing the 
  data to be monitored in Phase II and plotted on the foreground. Functions whose contribution to
  either the Hotelling's \(T^2\) and \(SPE\) statistics is out of
  control are plotted in red.
\end{itemize}

\begin{CodeChunk}
\begin{CodeInput}
R> plot_mon(cclist = cclist_pca, 
+           fd_train = mfdI[, c("X1", "X2", "X3")], 
+           fd_test = mfdII_pca[59])
\end{CodeInput}
\begin{figure}

{\centering \includegraphics[width=\textwidth]{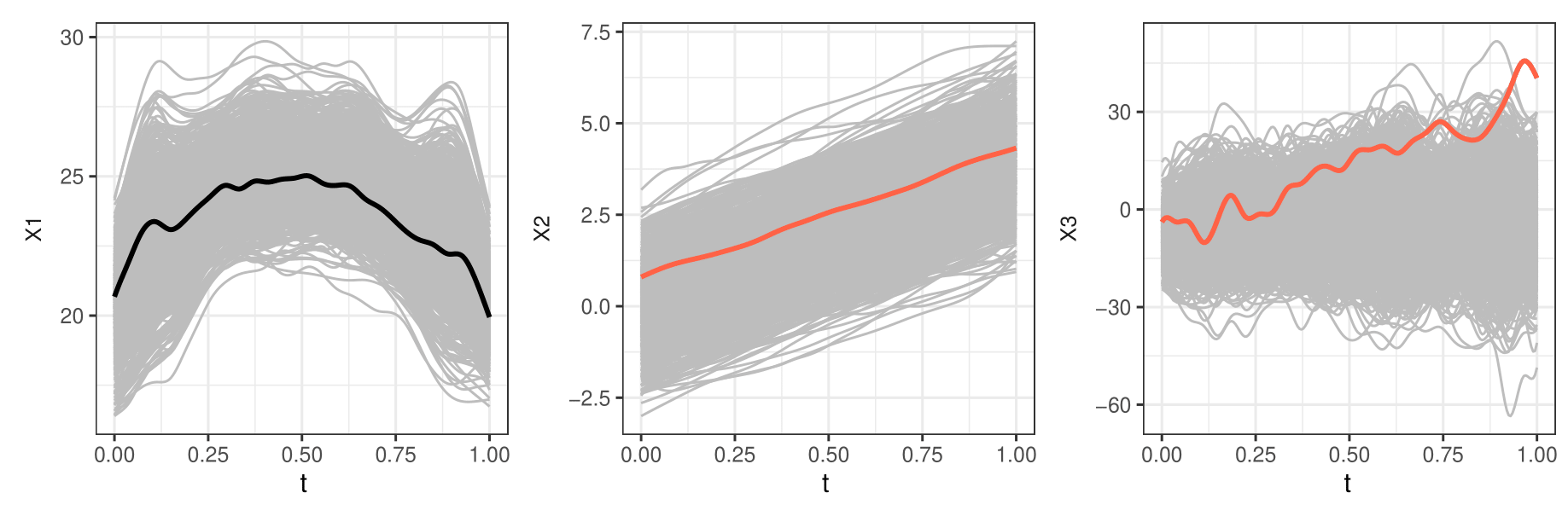} 

}

\caption{\label{fig:plot_mon_pca}Observation 59 of the functional covariates in the Phase II data set, plotted against the observations of the reference data set used for model fitting, plotted in gray. Red curves correspond to functional variables whose contribution to at least one monitoring statistic exceeds its limit, while the other curves are plotted as black lines.}
\end{figure}
\end{CodeChunk}

\subsection{Functional control charts for a univariate quality characteristic adjusted by the influence of covariates}
\label{sec_using_supervised}

The use of the \pkg{funcharts} package to implement functional control charts for a univariate quality characteristic adjusted by the influence of covariates  is illustrated in Sections \ref{sec_using_capezza} and \ref{sec_using_centofanti} with reference to scalar-on-function and function-on-function regression cases, respectively introduced in Sections \ref{sec_methods_capezza} and \ref{sec_methods_centofanti}.

\hypertarget{sec_using_capezza}{%
\subsubsection{Control charts based on scalar-on-function regression}\label{sec_using_capezza}}

To fit a scalar-on-function regression model with the reference data set,
the \pkg{funcharts} package provides the \texttt{sof\_pc()} function.
Its main arguments are:

\begin{itemize}
\tightlist
\item
  \texttt{y}, the vector containing the scalar response variable
  observations.
\item
  \texttt{mfdobj\_x}, an \texttt{mfd} object containing the
  observations of the functional covariates.
\item
  \texttt{tot\_variance\_explained},  the minimum desired fraction of variance
  to be explained by the set of PCs retained into the MFPCA model fitted on the
  functional covariates. Default value is 0.9.
\item
  \texttt{selection}, the criterion to choose the
  PCs to retain as predictors in regression model. 
  \texttt{"variance"} (default) retains the first \(M\) PCs such that together they explain a   fraction of variance greater than \texttt{tot\_variance\_explained},
  \texttt{"PRESS"} selects the \(j\)-th functional
  PC if, by adding it to the current set of selected PCs, the predicted residual error sum of squares (PRESS) statistic
  decreases. 
  \texttt{"gcv"} is similar to \texttt{"PRESS"}, where the PRESS statistic is substituted by the GCV score.
\end{itemize}

The following code uses default arguments and estimates the
scalar-on-function regression model with the reference data set:

\begin{CodeChunk}
\begin{CodeInput}
R> mod_sof <- sof_pc(y=d$datI$y_scalar, mfdobj_x=mfdI[,c("X1","X2","X3")])
\end{CodeInput}
\end{CodeChunk}

The output of \texttt{sof\_pc} is a list including the results of the regression of the scalar response on the scores, the results of the MFPCA obtained using \texttt{pca\_mfd()} and the estimated functional regression coefficient, as an object of class \texttt{mfd}.
The latter can be plotted using the \texttt{plot\_mfd()} function. 
Moreover, the \pkg{funcharts} package allows the graphical   uncertainty quantification  of the functional coefficient estimates via the \texttt{plot\_bootstrap\_sof\_pc()} function, which plots the coefficients estimated on the reference data set along with estimates obtained on bootstrap samples of the same data set, as in Figure \ref{fig:sof_beta}:

\begin{CodeChunk}
\begin{CodeInput}
R> plot_bootstrap_sof_pc(mod_sof, nboot = 100)
\end{CodeInput}
\begin{figure}

{\centering \includegraphics[width=\textwidth]{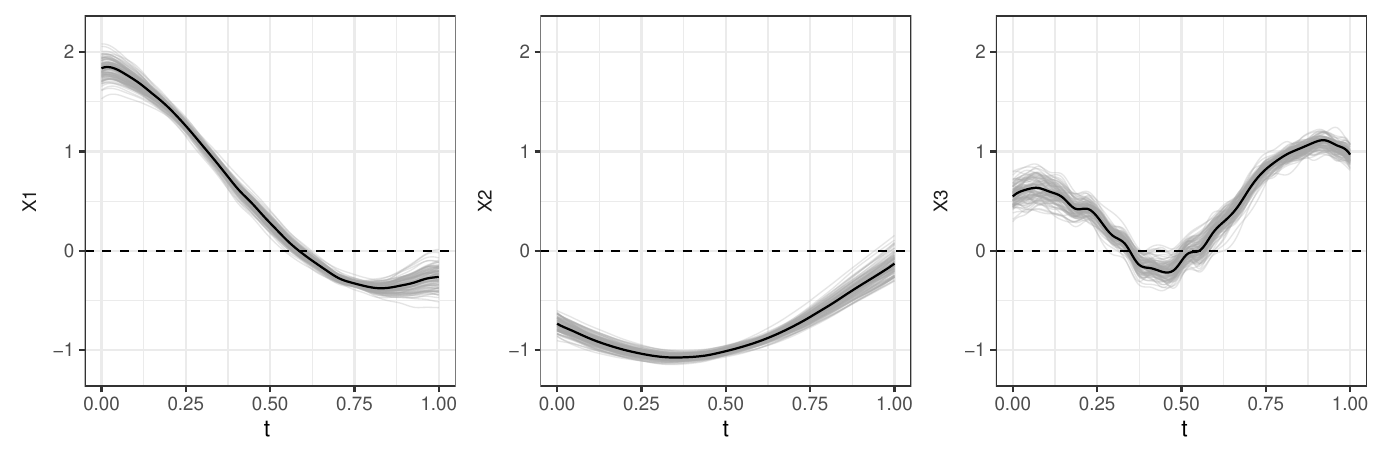} 

}

\caption{\label{fig:sof_beta}Estimated functional coefficients of the scalar-on-function regression model on the simulated reference data set, plotted in black, along with estimates of functional coefficients on 10 bootstrap samples, plotted in gray, for uncertainty quantification.}
\end{figure}
\end{CodeChunk}

Once the model has been estimated on the reference data set, it is possible to build the control charts and plot them for the prospective monitoring of the process, given the new Phase II data set. 
As stated in Section \ref{how_to_simulate}, the first group of data contains IC observations, in the second group of data, the scalar response has a moderate mean (\(d=1\)), while in the third group of data, the scalar response has a more severe mean shift (\(d=2\)).
Mean shifts of covariates are the same as in Section \ref{sec_using_colosimo}.
As explained in Section \ref{sec_methods_capezza}, \cite{capezza2020control} propose three control charts  based on the Hotelling's \(T^2\) and SPE to monitor the
multivariate functional covariates, together with the prediction error on the
scalar response variable to monitor the scalar response variable,
conditionally on the covariates. The \texttt{control\_charts\_sof\_pc()} function provides a data frame with all the
information required to plot the desired control charts. It has the
following main arguments:

\begin{itemize}
\tightlist
\item
  \texttt{mod}, the fitted model obtained as output of
  \texttt{sof\_pc}.
\item
  \texttt{mfdobj\_x\_tuning}, an optional data set of IC observations of the functional covariates to be used for control charts tuning.
  If no argument is provided, limits will be calculated as quantiles of the statistics computed on the reference data set.
\item
  \texttt{y\_test}, the vector containing the  observations of
  the scalar response variable to be monitored in Phase II.
\item
  \texttt{mfdobj\_x\_test}, the \texttt{mfd} object containing the
 observations of the multivariate functional covariates, corresponding to \texttt{y\_test}.
\item
  \texttt{alpha}, a list containing the Type I error for each control chart.
  The default value is \texttt{list(T2\ =\ 0.0125,\ spe\ =\ 0.0125,\ y\ =\ 0.025)}, which uses the Bonferroni correction to ensure that the overall Type I error \(\alpha\) is not   larger than \(0.05\).
\end{itemize}

The \texttt{plot\_control\_charts()} function can be used with the
obtained data frame to plot the three control charts as shown in Figure \ref{fig:sof_cc}.

\begin{CodeChunk}
\begin{CodeInput}
R> cclist_sof_pc <- control_charts_sof_pc(
+   mod = mod_sof,
+   mfdobj_x_tuning = mfdI_tun[, c("X1", "X2", "X3")],
+   y_test = d$datII_sof$y_scalar,
+   mfdobj_x_test = mfdII[, c("X1", "X2", "X3")])
R> plot_control_charts(cclist_sof_pc) &
+   geom_vline(aes(xintercept = c(20.5)), lty = 2) &
+   geom_vline(aes(xintercept = c(40.5)), lty = 2)
\end{CodeInput}
\begin{figure}

{\centering \includegraphics[width=\textwidth]{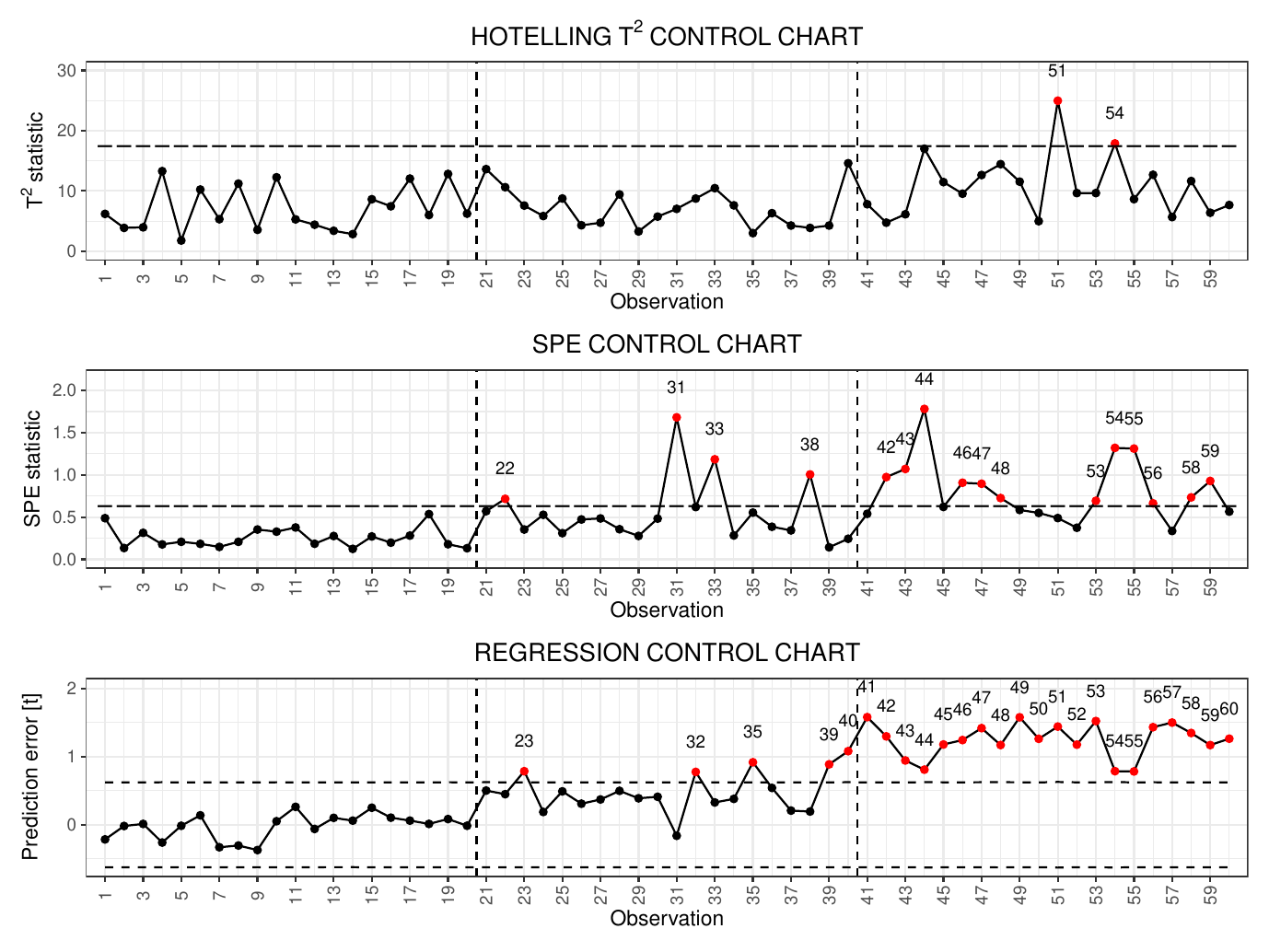} 

}

\caption{\label{fig:sof_cc}Functional control charts for the scalar-on-function regression case. Vertical lines separate the three groups of simulated Phase II data. Horizontal lines correspond to control chart limits, points correspond to the observations of the monitoring statistics for each Phase II observation. Red points highlights monitoring statistics that are out of the corresponding control limits; above each of them is reported the corresponding OC observation index.}
\end{figure}
\end{CodeChunk}

Since the multivariate functional covariates observations are the same as the data used in Section \ref{sec_using_colosimo}, by comparing Figures \ref{fig:sof_cc} and \ref{fig:pca_cc}, notice that the Hotelling's \(T^2\) and \(SPE\) control charts plot the same values of the monitoring statistics. 
The \texttt{control\_charts\_sof\_pc()} function calls in fact \texttt{control\_charts\_pca()} on the multivariate functional covariates. 
Control limits in Figure \ref{fig:sof_cc} are higher than those of Figure \ref{fig:pca_cc} because the Bonferroni correction takes into account the use of three control charts instead of two. 
The third control chart (for the scalar prediction error monitoring) signals 5 OC alarms over the second group of 20 data, while all 20 observations in the third group are above the upper control limit.
When an OC alarm is signaled by  the Hotelling's \(T^2\) or \(SPE\) control charts, one can  build contribution plots on the multivariate functional covariates as shown in Section \ref{sec_using_colosimo} and further visualize any multivariate functional observation against a reference data set to perform fault detection on multivariate functional covariates. 
When instead the response prediction error control chart issues an OC alarm, while the Hotelling's \(T^2\) and \(SPE\) charts do not, one concludes that possible anomalies may have occurred outside of the set of variables chosen as functional covariates.

\hypertarget{sec_using_centofanti}{%
\subsubsection{Control charts based on function-on-function regression}\label{sec_using_centofanti}}

To fit a function-on-function regression model with a reference data
set, the \pkg{funcharts} package provides the \texttt{fof\_pc()} function, having  arguments 

\begin{itemize}
\tightlist
\item
  \texttt{mfdobj\_y} and \texttt{mfdobj\_x}, the \texttt{mfd} objects with the Phase I observations of the functional response and covariates, respectively.
\item
  \texttt{tot\_variance\_explained\_x}, \texttt{tot\_variance\_explained\_y} and \texttt{tot\_variance\_explained\_res}, the minimum fraction of
  variance that have to be explained by the set of PCs retained into the MFPCA model fitted on the functional covariates, response and residuals, respectively. Default value is 0.95.
\item
  \texttt{type\_residuals}, the functional residual to be monitored (see choice (iii) of \cite{centofanti2020functional}). If \texttt{"standard"} (default), the
  MFPCA is calculated on the functional residuals. If
  \texttt{"studentized"}, the MFPCA  is calculated on the studentized
  residuals.
\end{itemize}

The following code uses default arguments and estimates the
function-on-function regression model with the reference data set:

\begin{CodeChunk}
\begin{CodeInput}
R> mod_fof <- fof_pc(mfdobj_y = mfdI[, "Y"],
+                    mfdobj_x = mfdI[, c("X1", "X2", "X3")])
\end{CodeInput}
\end{CodeChunk}

As output, \texttt{fof\_pc} returns a list including the results of the
regression of the functional response scores on the functional covariate
scores, the results of the MFPCA obtained using \texttt{pca\_mfd()} on the functional response,
the functional covariates and residuals. Moreover, the output list contains the
estimated  regression coefficient function, as an object of class
\texttt{bifd} from the \pkg{fda} package. Thus, it can be plotted via the
\texttt{plot\_bifd()} function provided by the \pkg{funcharts} package,
as in Figure \ref{fig:beta_fof}.

\begin{CodeChunk}
\begin{CodeInput}
R> plot_bifd(mod_fof$beta_fd)
\end{CodeInput}
\begin{figure}

{\centering \includegraphics[width=\textwidth]{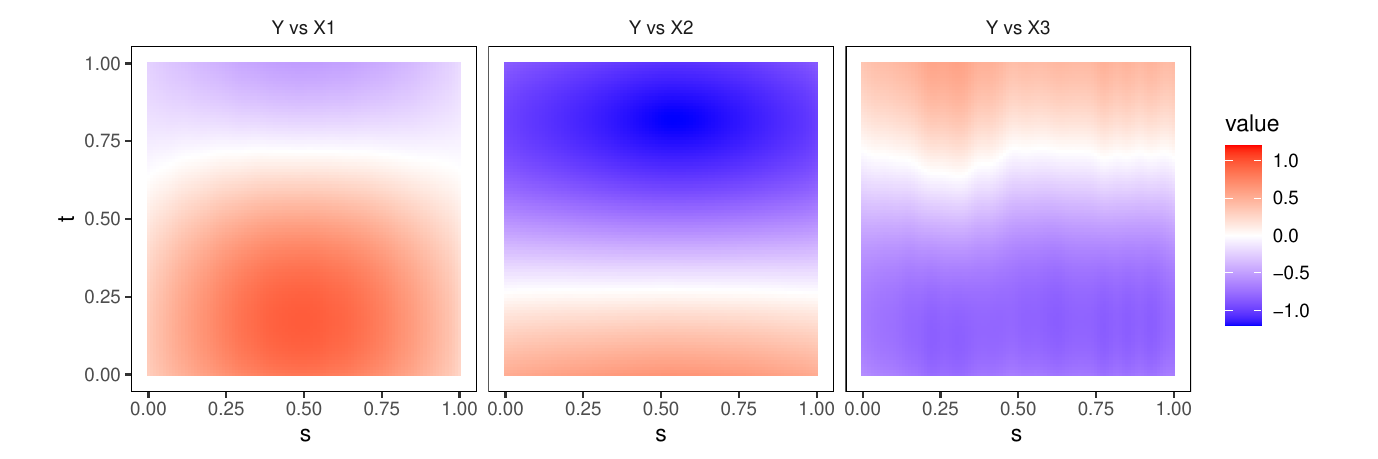} 

}

\caption{\label{fig:beta_fof}Plot of the the function-on-function regression coefficient estimated on the simulated reference data set.}
\end{figure}
\end{CodeChunk}

Once the model has been estimated on the reference data set, it is
possible to build the control charts for the prospective monitoring of the process, given the new Phase II data set.
As stated in Section \ref{how_to_simulate}, the first group of data contains IC observations, whereas, in the second and third groups of data, the functional response has a moderate mean shift of type D with severity \(d=0.5\) and \(d=1.5\), respectively.
The considered mean shifts of covariates are the same as in Section \ref{sec_using_colosimo}.
As explained in Section \ref{sec_methods_centofanti},  two control charts are used for
SPM in this case, based on the Hotelling's
\(T^2\) and the squared prediction error monitoring statistics,
calculated on the functional residuals of the function-on-function
regression model. The \texttt{regr\_cc\_fof()} function provides a data
frame with all the information required to plot the control chart. It has the following main arguments:

\begin{itemize}
\tightlist
\item
  \texttt{object}, the fitted model obtained as output of
  \texttt{fof\_pc}.
\item
  \texttt{mfdobj\_y\_tuning} and \texttt{mfdobj\_x\_tuning}, optional \texttt{mfd} objects of reference observations of the functional response and covariates, respectively, to be used for control chart tuning.
  If no argument is provided, limits will be calculated as quantiles of the statistics computed on the reference data set.
\item
  \texttt{mfdobj\_y\_new} and \texttt{mfdobj\_x\_new}, the \texttt{mfd} objects with the
   observations to be monitored in Phase II of the functional response and covariates, respectively.
\item
  \texttt{alpha}, the list containing the Type I error for the   Hotelling's \(T^2\) and the \(SPE\) control charts. 
  The default value is \texttt{list(T2\ =\ 0.025,\ spe\ =\ 0.025)}, which uses the Bonferroni correction to ensure  the overall Type I error \(\alpha\) is not   larger than \(0.05\).
\end{itemize}

The \texttt{plot\_control\_charts()} function can be used to plot
the control charts as in Figure \ref{fig:fof_cc}.

\begin{CodeChunk}
\begin{CodeInput}
R> cclist_fof_pc <- regr_cc_fof(
+   object = mod_fof,
+   mfdobj_y_tuning = mfdI_tun[, "Y"],
+   mfdobj_x_tuning = mfdI_tun[, c("X1", "X2", "X3")],
+   mfdobj_y_new = mfdII_fof[, "Y"],
+   mfdobj_x_new = mfdII_fof[, c("X1", "X2", "X3")])
R> plot_control_charts(cclist_fof_pc) &
+   geom_vline(aes(xintercept = c(20.5)), lty = 2) &
+   geom_vline(aes(xintercept = c(40.5)), lty = 2)
\end{CodeInput}
\begin{figure}

{\centering \includegraphics[width=\textwidth]{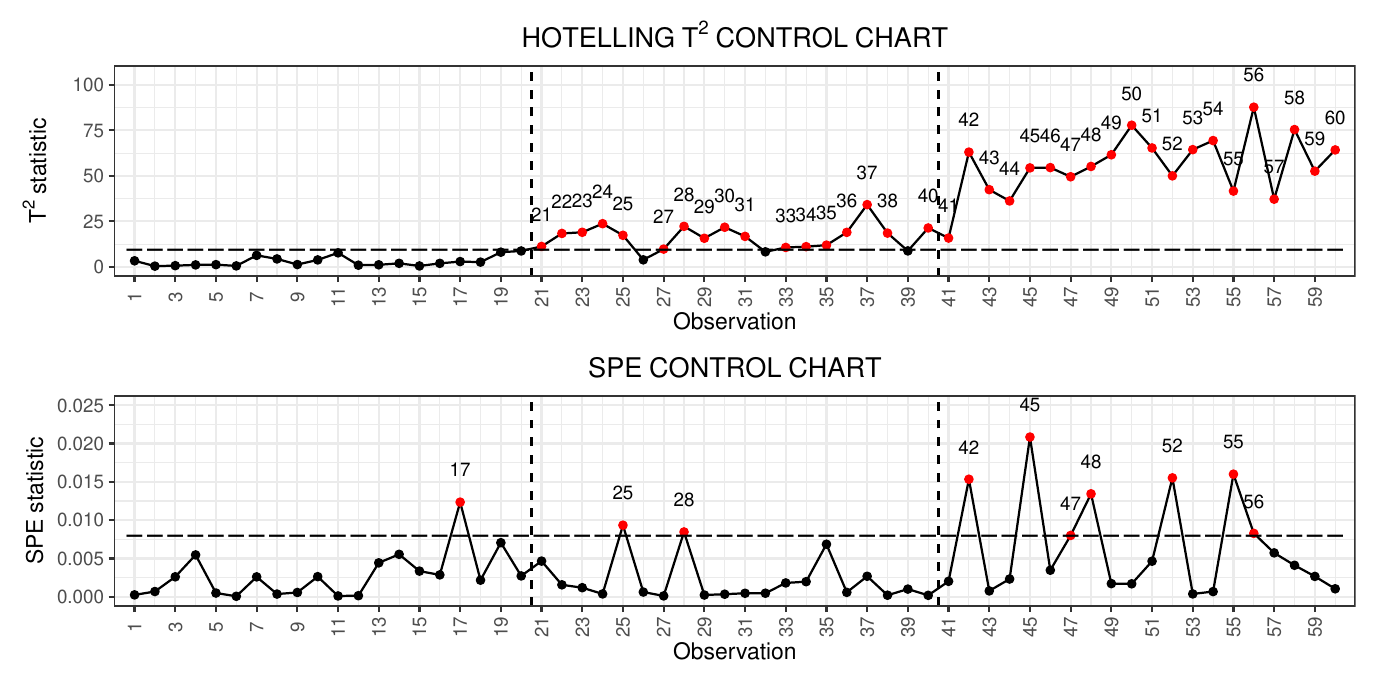} 

}

\caption{\label{fig:fof_cc}Phase II control charting for the function-on-function regression case. Vertical lines separate the three groups of simulated Phase II data. Horizontal lines correspond to control chart limits, points correspond to the observations of the monitoring statistics for each  observation. Red points and observation index highlight monitoring statistics that are out of the corresponding control limits.}
\end{figure}
\end{CodeChunk}

In the first group of  observations (generated from IC process) there is a false alarm signaled by the $SPE$ control chart, in the second group of observations (which are generated with moderate mean shift on both the third functional covariate and the functional response), only three observations within the control limits, while in the last group (with more severe mean shifts) all observations are signaled as OC.

Note that \texttt{regr\_cc\_fof()} provides control charts on the
prediction error of the functional response, without monitoring directly
the multivariate functional covariates. This means that, for example, an
observation with particularly anomalous values of some functional
covariates that still produces good prediction of the functional
response, such that the prediction error plots IC.
This happens because \cite{centofanti2020functional} focused on answering to the research question:
\textit{given the value of the functional covariates, is the observed response profile as expected?}

\hypertarget{sec_using_realtime}{%
\subsection{Real-time functional control charts}\label{sec_using_realtime}}

The \pkg{funcharts} package provides the real-time functional control charts stated in Section \ref{sec_methods_realtime} through the functions
\texttt{control\_charts\_pca\_mfd\_real\_time()},
\texttt{control\_charts\_sof\_pc\_real\_time()}, and
\texttt{regr\_cc\_fof\_real\_time()}, which are the real-time version of the main functions presented in Section \ref{sec_methods_colosimo}, \ref{sec_methods_capezza} and \ref{sec_methods_centofanti}, respectively. 
For shortness, in this section we illustrate only the real-time version of the functional control chart based on function-on-function regression introduced in Section \ref{sec_methods_centofanti} and implemented in Section \ref{sec_using_centofanti}.
Starting from the functional domain $\mathcal T = (a,b)$, in the following, we consider the evolution of the functional data over the subintervals $(a, a+k(b-a)))$, with $k$ varying along \texttt{k\_seq}, which is an additional input vector containing values defining the fraction of domain over which functional data are partially observed. 
By default, \texttt{k\_seq} is the sequence of values 0.2, 0.3, \dots, 1.
In the same way as the corresponding functions developed for the completely observed functional data, the \pkg{funcharts} package provides the functions \texttt{get\_mfd\_array\_real\_time()}, \texttt{get\_mfd\_list\_real\_time()} and \texttt{get\_mfd\_df\_real\_time()}, which accept as input lists of matrices, arrays and data frames in the long format. 
These functions return as output a list of \texttt{mfd} objects, each
partially observed on the interval $(a, a+k(b-a)))$ and create the data sets
required to perform the real-time functional control charts, as per the following code.

\begin{CodeChunk}
\begin{CodeInput}
R> xI <- get_mfd_list_real_time(d$datI[c("X1", "X2", "X3")])
R> yI <- get_mfd_list_real_time(d$datI["Y"])
R> xI_tun <- get_mfd_list_real_time(d$datI_tun[c("X1", "X2", "X3")])
R> yI_tun <- get_mfd_list_real_time(d$datI_tun["Y"])
R> xII <- get_mfd_list_real_time(d$datII[c("X1", "X2", "X3")])
R> yII <- get_mfd_list_real_time(d$datII["Y"])
\end{CodeInput}
\end{CodeChunk}

Then, the \texttt{fof\_pc\_real\_time()} function applies the \texttt{fof\_pc()} function to each element in the listis \texttt{yI} and
\texttt{xI}, and returns a list of equal length.

\begin{CodeChunk}
\begin{CodeInput}
R> mod_realtime <- fof_pc_real_time(yI, xI)
\end{CodeInput}
\end{CodeChunk}

The \texttt{regr\_cc\_fof\_real\_time()} function is used to apply \texttt{regr\_cc\_fof()} to each element in the list \texttt{mod\_realtime} and produce a data frame to plot the control charts for the real-time Phase II monitoring of  data sets \texttt{yII} and \texttt{xII}.

\begin{CodeChunk}
\begin{CodeInput}
R> cc_realtime <- regr_cc_fof_real_time(
+   mod_list = mod_realtime,
+   mfdobj_y_new_list = yII,
+   mfdobj_x_new_list = xII,
+   mfdobj_y_tuning_list = yI_tun,
+   mfdobj_x_tuning_list = xI_tun)
\end{CodeInput}
\end{CodeChunk}

We can finally plot the real-time functional control charts for a single
observation, which shows, for each \(k\), the monitoring statistics
calculated on the data observed in $(a, a+k(b-a)))$. 
Figure \ref{fig:realtime_cc} shows an example for the future observation 30, belonging to the second group of Phase II observations (see Section \ref{how_to_simulate}). 
Because of the mean shift in both the third covariate and the response, an anomalous behaviour is signalled  in the final part of the functional domain by the Hotelling's \(T^2\).
Also note in Figure \ref{fig:realtime_cc} that the monitoring statistics corresponding to $k=1$, i.e., when the profiles are fully observed, are equal to the monitoring statistics of the future observation 30 shown on the control chart in Figure \ref{fig:fof_cc}.

\begin{CodeChunk}
\begin{CodeInput}
R> plot_control_charts_real_time(cc_realtime, id_num = 30)
\end{CodeInput}
\begin{figure}

{\centering \includegraphics[width=.6666667\textwidth]{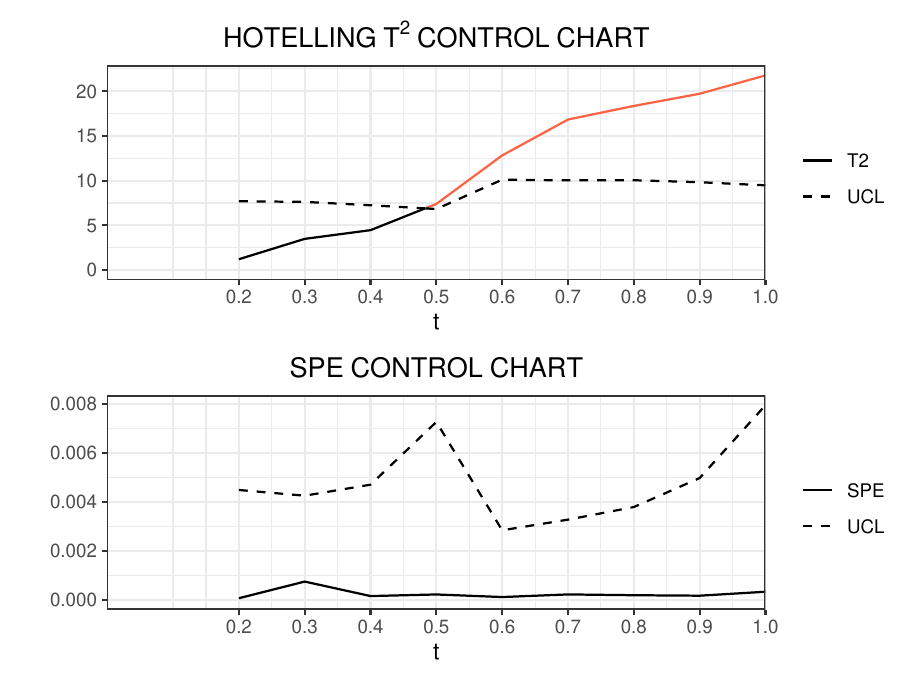} 

}

\caption{\label{fig:realtime_cc}Real-time control charts based on function-on-function regression to monitor the simulated Phase II observation 30. Solid lines show the profiles of the monitoring statistics (T2 and SPE), coloured in red if they are above the corresponding upper control limits (UCL), drawn as dashed lines.}
\end{figure}
\end{CodeChunk}

\section{A real-case study: ship CO$_{2}$ emission monitoring}

\label{sec_realcasestudy}

To illustrate the use of the \pkg{funcharts} package in practical cases, we present in this section a real-case study on the SPM of Ro-Pax ship CO\textsubscript{2} emissions during navigation, which is defined as the time interval between the finished with engine order (when the ship leaves the departure port) and the stand by engine order (when the ship enters the arrival port). 
The topic has become extremely relevant in the recent years in view of the dramatic climate change. At the European level, the Regulation EU 2015/757 of the European Union (EU) Council, by following the guidelines of the International Maritime Organization, has urged ships operating in the Mediterranean Sea to adopt systems for monitoring, reporting and verification  of CO\textsubscript{2} emissions. 
Even though only summary statistics are strictly required by this regulation, shipping companies are equipping their fleet with multi-sensor systems that make available large amounts of high-frequency operational data, which can be naturally regarded as functional data. 
In this real-case study, we use a part of the \texttt{ShipNavigation} data set, which is courtesy of the multinational logistic group and shipping company Grimaldi Group and is available in the Supplementary Material. 
The complete data set contains four years' worth of data collected in real time with five-minute frequency on board of a Ro-Pax ship connecting three main ports of the Mediterranean sea, even though some minor ports are rarely used as intermediate steps.

For confidentiality reason, the name of the ship and ports as well as the gps data are omitted. 
The \texttt{datetime} variable counts the number of seconds from a reference omitted date. 
Most of the variables are scaled by dividing  by the maximum absolute value. 
In particular, \texttt{CO2\_emissions} are measured indirectly based on the fuel consumption, which in turn is calculated based on the engine power measurements.
Each variable profile refers to a ship voyage connecting a departure and an arrival port and is defined as a function of the fraction of the total navigation distance traveled by the ship at each voyage, i.e., on the domain  \(\mathcal T = (0, 1)\).
Voyages of the complete data set are identified by a unique voyage number (VN) label with a progressive number and is stored in the \texttt{VN} column, so that time
order is preserved. 
Moreover, it is known that an energy efficiency initiative (EEI) was performed on the ship between VN1251 and VN1252 and plausibly produced a mean shift in the CO\textsubscript{2} emissions. 
For more details on the variable description the reader may  refer to \cite{erto2015procedure,bocchetti2015statistical}.

In this real-case study, we focus on the Phase II monitoring of CO\textsubscript{2} emissions after the EEI during navigation on a specific route from port C to port A, through profiles of the CO\textsubscript{2} emissions per mile (identified by the variable \texttt{CO2pm}) adjusted by the influence of four functional covariates, namely the ship speed over ground \texttt{sog} and the \texttt{trim}, as well as the longitudinal (\texttt{w\_long}) and transverse (\texttt{w\_trasv}) components of the wind.
Thus, we can apply the methods described in Section \ref{sec_methods_centofanti}.
If instead the interest is in a scalar response, such as the total CO\textsubscript{2} emissions at each voyage, the reader may want to refer to the methods described in Section \ref{sec_methods_capezza}.
The reference data set of this real-case study is then obtained by including one year's worth of data  up to the last voyage before the EEI pertaining to the considered route, i.e., the 159
voyages between VN0519 and VN1248. 
Then, Phase II monitoring starts from VN1257.

In the Supplementary Material S3 we provide the pre-processing code that loads the \texttt{ShipNavigation} data set and returns the \texttt{mfd} objects \texttt{y1}, \texttt{x1}, \texttt{y2} and \texttt{x2}, which contain the observations of the functional response and covariates, divided  into reference data set and future observations to be used in Phase II. 
Given these data, the function-on-function regression model can be fit as follows and the functional control chart based on studentized residuals is built.

\begin{CodeChunk}
\begin{CodeInput}
R> mod <- fof_pc(y1, x1, type_residuals = "studentized")
R> plot_bifd(mod$beta_fd)
R> cc <- regr_cc_fof(mod, mfdobj_y_new = y2, mfdobj_x_new = x2,
+                    alpha = list(T2 = 0.005, spe = 0.005))
R> cutpoint <- which(c(obs1, obs2) > "VN1248")[1]
R> plot_control_charts(cc) & geom_vline(aes(xintercept=cutpoint), lty=2)
\end{CodeInput}
\begin{figure}

{\centering \includegraphics[width=\textwidth]{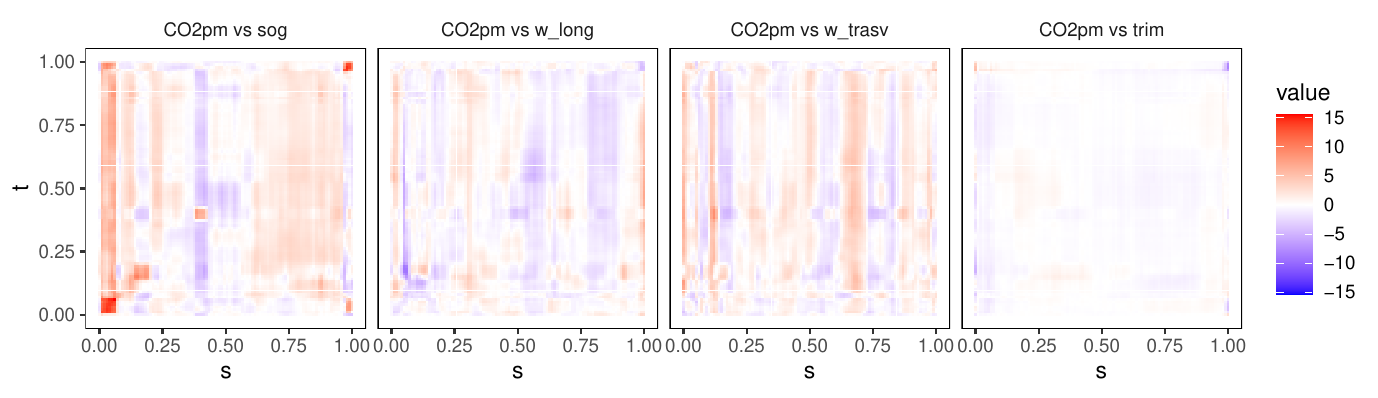} 

}
\caption{\label{fig:beta_fof_real}Estimated functional coefficient of the function-on-function regression model on the Phase I navigation data set.}
\end{figure}
\begin{figure}

\centering \includegraphics[width=\textwidth]{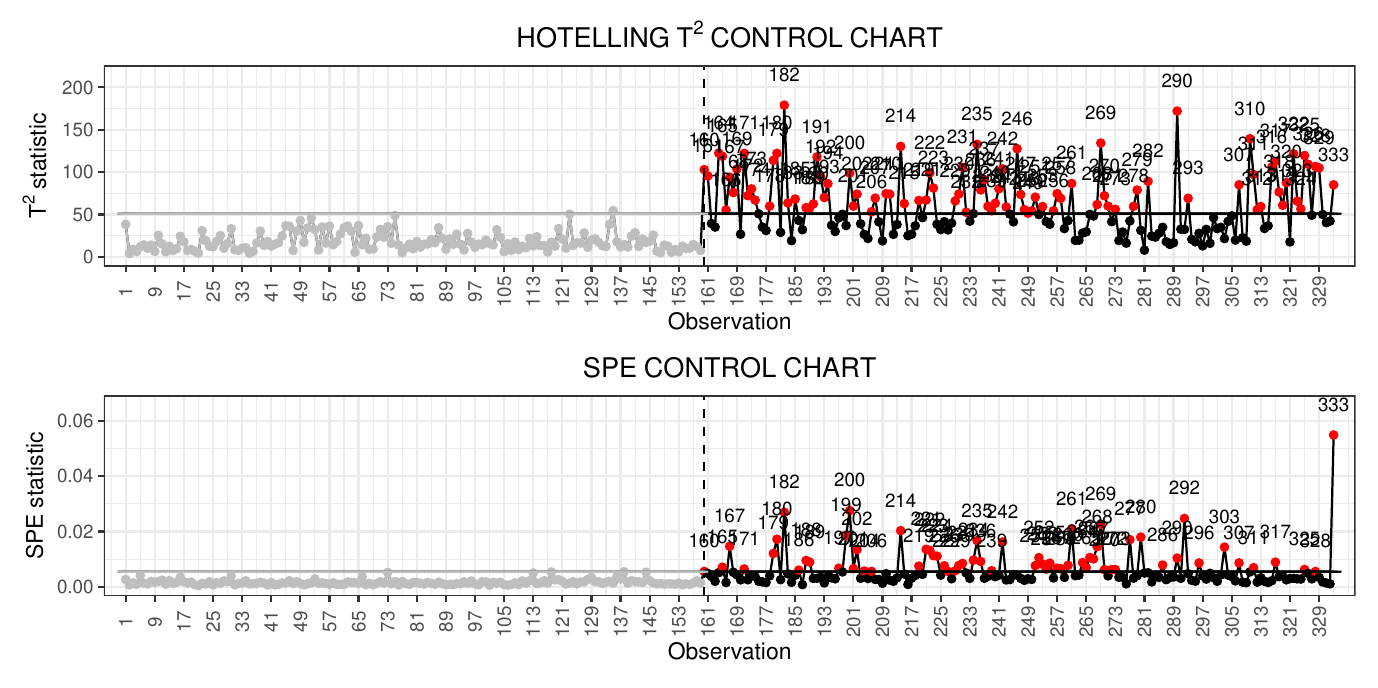} 
\caption{\label{fig:cc_real}Functional control chart for the real-case study on the \texttt{ShipNavigation} data. The vertical dashed line separates the reference data set, whose points are reported in gray, from the Phase II data set. Horizontal lines correspond to control chart limits, points correspond to the observations of the monitoring statistics for each observation. Red points and observation index highlight monitoring statistics that plot outside of the corresponding control limits.}
\end{figure}
\end{CodeChunk}

Figure \ref{fig:beta_fof_real} shows the estimated functional regression coefficients, while Figure \ref{fig:cc_real} shows the Phase II control charts. 
Note that  the monitoring statistics calculated on the reference data set are plotted, for  convenience of graphical comparison, on the left hand side of the vertical dashed line in gray. 
We set an overall Type I error \(\alpha = 0.01\) and use the Bonferroni correction to take into account the simultaneous use of the two control charts.
As expected, both Hotelling's $T^2$ and SPE control charts show a clear mean shift in the functional studentized residuals after EEI, which is marked by a dashed vertical line. 

To illustrate the real-time application of these control charts, we consider the last observation shown in Figure \ref{fig:cc_real}, that is observation 333 (identified by VN2054).
The real-time Hotelling's $T^2$ and $SPE$ control charts for this observation, plotted in Figure \ref{fig:real_time_frcc_real}, are obtained through the following code.
\begin{CodeChunk}
\begin{CodeInput}
R> modl <- fof_pc_real_time(mfdobj_y_list = y1l,
+                           mfdobj_x_list = x1l,
+                           type_residuals = "studentized")
R> ccl <- regr_cc_fof_real_time(mod_list = modl,
+                               mfdobj_y_new_list = y2l,
+                               mfdobj_x_new_list = x2l,
+                               alpha = list(T2 = 0.005, spe = 0.005))
R> plot_control_charts_real_time(ccl, 333)
\end{CodeInput}
\begin{figure}

\centering \includegraphics[width=.666667\textwidth]{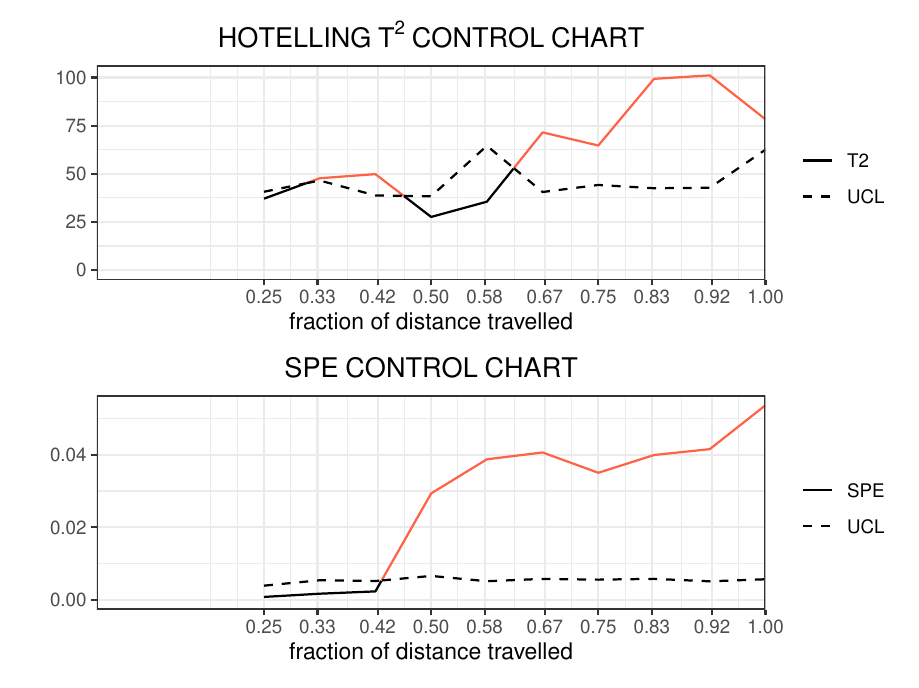} 

\caption{\label{fig:real_time_frcc_real}Real-time functional control chart to monitor the Phase II observation 333 in the navigation data set. Solid lines show the profiles of the monitoring statistics (T2 and SPE), coloured in red if they are above the corresponding upper control limits (UCL), drawn as dashed lines.}
\end{figure}
\end{CodeChunk}
The lists of \texttt{mfd} objects \texttt{y1l}, \texttt{x1l}, \texttt{y2l} and \texttt{x2l}, containing the real-time observations of the functional response and covariates are obtained through the pre-processing code provided by Supplementary Material S3.
Figure \ref{fig:real_time_frcc_real} shows the real-time functional control chart plotted for the observation 333. 
We can notice that the monitoring statistics go very large in the central part of the domain, where the functional residual profile is shown to be particularly low (Figure \ref{fig:oc_voyage}).

Once the voyage is completed, the \texttt{predict\_fof\_pc()} function can be used to obtain the functional studentized residuals of a given voyage, e.g., the  OC profile of the studentized residual corresponding to observation 333.
Using the function \texttt{plot\_mon}, in Figure \ref{fig:oc_voyage} this profile is superimposed on the functional studentized residual profiles calculated on the reference data set and stored in the object \texttt{mod} by the following code.

\begin{CodeChunk}
\begin{CodeInput}
R> yhat <- predict_fof_pc(object=mod,mfdobj_y_new=y2,mfdobj_x_new=x2)
R> plot_mon(cc, mod$residuals, yhat$pred_error[333])
\end{CodeInput}
\begin{figure}

{\centering \includegraphics[width=0.333333\textwidth]{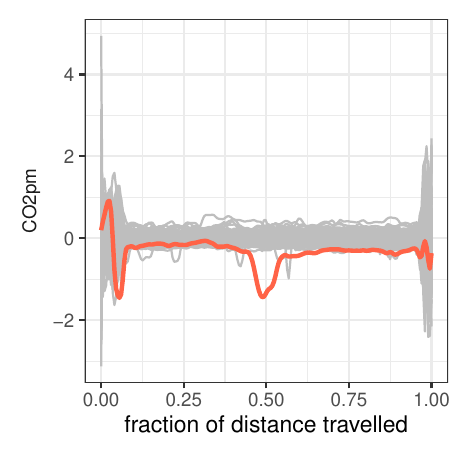} 

}

\caption{\label{fig:oc_voyage} The  OC profile of the studentized residual corresponding to the observation 333, superimposed in red on the  functional studentized residual profiles calculated on the reference data set.}
\end{figure}
\end{CodeChunk}

In accordance with Figure \ref{fig:real_time_frcc_real}, Figure \ref{fig:oc_voyage} shows a large negative prediction error for CO\textsubscript{2} emissions of observation 333 in the central part of
the voyage, which have led to large values of the monitoring statistics.
This highlights that, conditionally on the functional covariates, in
that part of the voyage, the profile of the CO\textsubscript{2} emissions
per mile is particularly lower than expected.

\section{Conclusions}
\label{sec_conclusions}

This article illustrates the \pkg{funcharts} R package, which is the first off-the-shelf toolkit for the SPM of processes characterized by quality characteristics in the form of functional data.
In particular, \pkg{funcharts} implements the methods proposed by 
\cite{colosimo2010comparison} and the recent SPM frameworks of
\cite{capezza2020control} and \cite{centofanti2020functional}. 
The package provides practitioners with tools to generate functional data, to perform multivariate functional principal component analysis, to build control
charts for the monitoring of shifts in the process function mean. 
The real-time version of these tools is also provided when functional data are observed only up to a given point. 
We demonstrate the package through its built-in data generator as well as a real-case study on the monitoring of CO\textsubscript{2} emissions during the navigation of a Ro-Pax ship, courtesy of the Italian shipping company Grimaldi Group.


\section{Computational details}
\label{sec_comp_details}

The results in this article are obtained using 64-bit R 4.1.2 with the \pkg{funcharts} 1.2.0 package and a computer with an
Intel\textsuperscript{\textcopyright} Core\textsuperscript{TM} i9-9980HK CPU at 2.40GHz 32 GB of RAM on Ubuntu 20.04. This package makes use of
the \pkg{fda} 5.5.1 package to get functional data from discrete
observations and to fit the multivariate functional principal component.
 The package depends on the \pkg{ggplot2} 3.3.5 and
\pkg{patchwork} 1.1.1 packages, for plotting functional data and control
charts, and   \pkg{parallel} 4.1.2 package, for
the parallel computation. 
\proglang{R} itself and these packages are
available from the Comprehensive R Archive Network (CRAN) at
\url{https://CRAN.R-project.org/}.
On this computer, the computation time required to produce all the results shown in Section \ref{sec_using} was about 90 seconds, while the real case-study in Section \ref{sec_realcasestudy} required about 25 minutes of computation time.

\section{About the authors}
\label{about-the-authors}

\textbf{Christian Capezza} is a researcher in Statistics for experimental
and technological research at the Department of Industrial Engineering of the University of Naples Federico II. 
He works on advanced statistical methodologies for engineering applications and his research project regards the development of interpretable
statistical methods for the analysis of complex systems in Industry 4.0.

\textbf{Fabio Centofanti} is a post-doc researcher at the Department of Civil, Building and Enviromental Engineering of the University of Naples Federico II, Italy.
His main research interests include functional data analysis and statistical process monitoring for industrial applications.

\textbf{Antonio Lepore} is an Associate Professor of Statistics for experimental and technological research at the Department of Industrial
Engineering of the University of Naples Federico II, Italy. His main
research interests include the industrial application of statistical
techniques to the monitoring of complex measurement profiles from
multisensor acquisition systems, with particular attention to renewable
energy and harmful emissions.

\textbf{Alessandra Menafoglio} is an Assistant Professor at MOX, Department of
Mathematics, Politecnico di Milano. Her research interests focus on the
development of innovative statistical models and methods for the
analysis and statistical process control of complex observations
(e.g.~curves, images and functional signals), possibly characterized by
spatial dependence.

\textbf{Biagio Palumbo} is an Associate Professor of Statistics for experimental
and technological research at the Department of Industrial Engineering
of the University of Naples Federico II, Italy. His major research
interests include reliability, design and analysis of experiments,
statistical methods for process monitoring and optimization and data
science for technology.

\textbf{Simone Vantini} is an Associate Professor of Statistics at the
Politecnico di Milano, Italy. He has been publishing widely in
Functional and Object-Oriented Data Analysis. His current research
interests include permutation testing, non-parametric forecasting,
process control, non-Euclidean data and in general statistical methods
and applications motivated by business or industrial problems.

\section{Acknowledgements}
\label{acknowledgements}
The authors are extremely grateful to the Grimaldi Group’s Energy Saving Department engineers Dario Bocchetti and Andrea D’Ambra for the access to the \texttt{ShipNavigation} data and the general support over the course of these activities.

\section{Funding}
The present work was partly developed within the activities of the project ``Functional Data Analysis for Traffic Operations'' funded by ``Programma per il finanziamento della ricerca di Ateneo – Linea B'' of the University of Naples Federico II (ref. ALTRI\_CdA\_75\_2021\_FRA\_LINEA\_B).

\renewcommand\refname{References}
\bibliography{refs.bib}

\end{document}


\title{Supplementary Material: \\funcharts: Control charts for multivariate functional data in
\proglang{R}}

\date{}

\author[]{}

\renewcommand{\theequation}{S\arabic{equation}}
\renewcommand{\thesection}{S\arabic{section}}

\maketitle

\section{Details on simulation}
\label{appendix_sim}

\subsection{Mathematical details}

For the control charts based on scalar-on-function regression, described in Section 2.3.1, mathematical details on how data are generated are given as follows.
To generate the scalar response, we need to define the regression coefficients in Equation (5).
The scalar intercept is set as $\beta_0=0$, while, as shown in Equation (6), after basis expansion functional coefficients we need to set coefficients $b_1, \dots, b_{M}$, where we set $M=50$, depending on the desired value of $R^2$, where $R^2=\text{var}(E(y|X_1 (s), \dots, X_P (s)) / \text{var}(y)$.
We consider three values for $R^2$.
In particular, for $l=1, \dots, 10$, to achieve $R^2$ = 0.97 we set $b_l = 0.587709$, while to achieve $R^2$ = 0.86 we set $b_l = 0.5533828$, and to achieve $R^2$ = 0.74 we set $b_l = 0.5133249$. 
Note that we set $b_{l} = 0$ for $l > 10$.
Then, in order to ensure the validity of the model in Equation (5), realizations of the scalar response variable are obtained as $y_i = \sum_{m=1}^{M} \xi_{im} b_m + \varepsilon_i$, where $\xi_{im}$ are the scores of the multivariate functional covariates and we generate $\varepsilon_i \sim N(0, \sigma^2_{\varepsilon})$ independent of the functional covariates, with $\sigma^2_{\varepsilon} = \sigma^2 - \sum_{m=1}^{M} b_m^2 \lambda_m$, where $\lambda_m$ are the variances of the scores. 
Finally, a mean shift on the scalar response is obtained by adding to the simulated values a fixed quantity equal to the severity $d$.
For mathematical details on the function-on-function regression case, the reader is referred to \cite{centofanti2020functional}.

\subsection{Implementation details}

The function \texttt{sim\_funcharts()} relies on the function \texttt{simulate\_mfd()} to simulate functional data with \pkg{funcharts}.
The function \texttt{simulate\_mfd()}
generates a data set with three functional covariates, a functional
response generated as a function of the three functional covariates
through a function-on-function linear model, and a scalar response
(independent of the functional one) generated as a function of the three
functional covariates through a scalar-on-function linear model. 

In particular, each simulated observation of a functional variable consists
of a vector of 150 discrete points, equally spaced on the functional
domain \(\mathcal{T} = (0, 1)\), generated with noise. In the default
case, the function generates data from the process when it is in control. Additional arguments can
be used to generate data from an out-of-control process, with mean shifts
according to the scenarios proposed by \cite{centofanti2020functional}.
Since functional data are observed on a regular grid, the function
returns a list of matrices, each corresponding to a functional
variable, and a vector for the scalar response. The matrices have the
number of rows equal to the sample size and the number of columns equal
to 150, i.e., the number of discrete points of each functional
observation.

The main arguments of \texttt{simulate\_mfd()} are:

\begin{itemize}
\tightlist
\item
  \texttt{nobs}: the number of observations to be generated.
\item
  \texttt{R2}: the \(R^2\) of the regression model.
\item
  \texttt{shift\_type\_y}, \texttt{shift\_type\_x1},
  \texttt{shift\_type\_x2}, \texttt{shift\_type\_x3}: the shift type on
  the mean of the corresponding functional variable, according to
  Appendix \ref{appendix_sim}.
\item
  \texttt{d\_y}, \texttt{d\_x1}, \texttt{d\_x2}, \texttt{d\_x3},
  \texttt{d\_y\_scalar}: the severity of the mean shift of the
  corresponding variable.
\end{itemize}

In the following, we report the code of the function \texttt{sim\_funcharts()}:

\begin{verbatim}
R> datI <- simulate_mfd(nobs=1000)
R> datI_tun <- simulate_mfd(1000)
R> datII_ic <- simulate_mfd(20)
R> datII_oc1 <- simulate_mfd(20, shift_type_x3="A", d_x3=20,
+                            d_y_scalar=1, shift_type_y="D", d_y=0.5)
R> datII_oc2 <- simulate_mfd(20, shift_type_x3="A", d_x3=40,
+                            d_y_scalar=2, shift_type_y="D", d_y=1.5)
R> datII <- list()
R> for (ii in 1:4) {
+    datII[[ii]] <- rbind(datII_ic[[ii]],datII_oc1[[ii]],datII_oc2[[ii]])
+    datII[[ii]] <- rbind(datII_ic[[ii]],datII_oc1[[ii]],datII_oc2[[ii]])
+    datII[[ii]] <- rbind(datII_ic[[ii]],datII_oc1[[ii]],datII_oc2[[ii]])
+  }
R> datII[[5]] <- c(datII_ic[[5]],datII_oc1[[5]],datII_oc2[[5]])
R> names(datII) <- names(datI)
R> return(list(datI=datI, datI_tun=datI_tun, datII=datII))

\end{verbatim}

\section{Details on the mfd class}
\label{appendix_mfd}

The \pkg{funcharts} package provides the \texttt{mfd} class for
multivariate functional data, which inherits from the \texttt{fd} class of the \pkg{fda} package and provides some additional features. 
In particular, it forces the \texttt{coef} argument to be a
three-dimensional array even when dealing with single functional
observations and/or variables and it provides a better subset function
\texttt{{[}} that never drops dimensions, i.e., it always returns a
\texttt{mfd} object with three-dimensional array argument, moreover it
allows extracting observations/variables also by name; when possible, it
stores the original raw data in the long data frame format, which is useful for plotting the raw discrete data instead of the smoothed
profiles with the functions provided by the package.

The first step in functional data analysis is usually to get smooth
functional data from the original discrete observations. The
\pkg{funcharts} package obtains smooth functional data as described in Section 2.1, where in particular the smoothing parameter is
selected by minimizing the generalized cross-validation criterion over a
grid of values. To this purpose, the \pkg{funcharts} package provides
several functions depending on the data type passed as input:

\begin{itemize}
\tightlist
\item
  \texttt{get\_mfd\_df()}: for a data frame in the long format, with one
  column whose name must be passed as a string to the \texttt{arg}
  argument, giving the domain values, one column whose
  name must be passed as a string to the \texttt{id} argument, denoting
  a categorical variable indicating the functional observation, and one
  column per each functional variable indicating the corresponding
  \texttt{y} values. Names of functional variables must be passed as a
  vector of strings to the \texttt{variables} argument
\item
  \texttt{get\_mfd\_list()}: for a list of matrices, in the case all
  functional data are observed on the same grid, where, for each matrix,
  each row corresponds to a functional observation and each column
  corresponds to a point on the grid.
\item
  \texttt{get\_mfd\_array()}: for a three-dimensional array, in the case
  all functional data are observed on the same grid, where the first
  dimension corresponds to grid points, the second to replications, and
  the third to variables within replications.
\end{itemize}

The main arguments of these functions are:

\begin{itemize}
\tightlist
\item
  \texttt{n\_basis}: the number of B-spline basis functions, default is
  30.
\item
  \texttt{lambda\_grid}: a grid of values from which the smoothing
  parameters have to be chosen for each functional datum.
\end{itemize}

In the \pkg{funcharts} package, we assume functional data are densely
observed over their compact domain. However, it is important to
distinguish the two cases where all functional data are observed on a
common grid, from the one where each functional observation has a its
own set of grid points. In \pkg{funcharts}, the former case is covered
by the functions \texttt{get\_mfd\_array()} and
\texttt{get\_mfd\_list()}, while the latter requires to use the function
\texttt{get\_mfd\_df()} with a data frame in the long format, as in the
real-case study shown in Section 4. From a
computational point of view, having different grid points for each
functional observation requires longer calculations. Let us focus on the
\(i\)-th sample of a single functional variable \(y_i(t)\) observed on a
grid \(t_{i1}, \dots, t_{ig_i}\) of \(g_i\) domain points. Let us
estimate the functional observation given a fixed smoothing parameter
\(\lambda\). If we have \(K\) B-spline basis functions
\(B_1(t), \dots, B_K(t)\), we can denote the \(g_i \times K\) design
matrix containing the basis function evaluations as
\(Z_i=(z_{i1}, \dots, z_{iK})\), where
\(z_{ik}=(B_k(t_{i1}), \dots, B_k(t_{ig_i}))^\top\) and the vector of
the observed functional values as
\(y_i=(y_i(t_{i1}), \dots, y_i(t_{ig_i}))^\top\). If we denote the
\(K \times K\) matrix penalizing the integrated squared second
derivatives as \(S\) whose \((i,j)\)-th element is
\(S_{i,j}=\int_{\mathcal T} B_{i}^{(2)}(t) B_{j}^{(2)}(t) dt\), then we
can estimate the function as \(y_i(t) = \sum_{k=1}^K b_{ik} B_{k}(t)\),
where
\(b_i = (b_{i1}, \dots, b_{iK})^\top = (Z_i^\top Z_i + \lambda S)^{-1} Z_i^\top y_i\).
If the grids containing the domain points over which functional data are
observed are different, we require as many matrix inversions as the
sample size \(n\) to get all functional data for a fixed smoothing
parameter. If instead all functional data are observed on the same grid
\(t_1, \dots, t_g\), then all \(i\) subscripts above disappear and we
can arrange all functional values into a \(g \times n\) matrix
\(Y = (y_1,\dots,y_n)\), where \(y_i=(y_i(t_1), \dots, y_i(t_g))\). This
means that with a single matrix inversion we get all basis coefficients
together in the \(K \times n\) matrix \(B = (b_{1}, \dots, b_{n})\),
where \(b_i=(b_{i1}, \dots, b_{iK})^\top\), i.e.,
\(B=(Z^\top Z + \lambda S)^{-1} Z^\top Y\). In order to save
computational time with the long data frame case through
\texttt{get\_mfd\_df()}, we can use parallel computation by setting the
argument \texttt{ncores} greater than 1.

\section{Pre-processing of the \texttt{ShipNavigation} data set}
\label{appendix_grimaldi}

In this Section, we show how the \texttt{ShipNavigation} data set, provided as Supplementary Material, is loaded and pre-processed in order to obtain the data analysed in Section 4.
First, let us load the data, then we need to calculate the
CO\textsubscript{2} emissions per mile and filter only the required
rows, corresponding to the considered route from port C to port A.

\begin{verbatim}
R> load("ShipNavigation.RData")
R> dft <- ShipNavigation %>%
+   group_by(VN) %>%
+   mutate(CO2pm = CO2_emissions / nautic_miles) %>%
+   ungroup() %>% 
+   filter(position == "C -> A") %>% 
+   na.omit()
\end{verbatim}

Now we can use the \pkg{funcharts} package to get functional data from
the original observations. In this application we choose 100 B-spline
basis functions. Note that from the Phase I analysis, which we do not
include here for brevity and because the focus of \pkg{funcharts} is on
Phase II monitoring, we remove two outlier voyages, VN0920 and VN0533.
The following code builds the \texttt{mfd} object that will be used in
the analysis, it removes the outliers, and it creates the functional
response and covariates \texttt{mfd} objects for Phase I and II,
i.e.~\texttt{y1}, \texttt{x1}, \texttt{y2}, \texttt{x2}.

\begin{verbatim}
R> yvar <- "CO2pm"
R> xvar <- c("sog", "w_long", "w_trasv", "trim")
R> 
R> x <- get_mfd_df(dt = dft,
+                  arg = "percent_miles",
+                  domain = c(0, 1),
+                  id = "VN",
+                  variables = c(xvar, yvar),
+                  n_basis = 100)
R> outliers <- c("VN0920", "VN0533")
R> obs <- unique(dft$VN)
R> obs <- obs[!(obs %in% outliers)]
R> x <- x[obs]
R> obs1 <- obs[obs <= "VN1249" & obs >= "VN0517"]
R> obs2 <- obs[obs > "VN1249" & obs <= "VN2054"]
R> x1 <- x[obs1, xvar]
R> y1 <- x[obs1, yvar]
R> x2 <- x[c(obs1, obs2), xvar]
R> y2 <- x[c(obs1, obs2), yvar]
\end{verbatim}

Note that in \texttt{y2}, \texttt{x2} we include both observations \texttt{obs1} and \texttt{obs2} are included.
Therefore, control charts using these objects will show both Phase I and Phase II data.

For the implementation of the real-time functional control charts, the registration of functional data, which is a problem beyond the scope of the \pkg{funcharts} package, has already been applied in advance, then the \texttt{ShipNavigation} data set already includes the column \texttt{percent\_miles\_hat} that estimates the fraction of distance traveled in real time as proposed by \cite{capezza2020control}. 
Note that the procedure has only been applied on the routes with at least 10 voyages.


The following code builds the lists of mfd objects \texttt{y1l}, \texttt{x1l}, \texttt{y2l} and \texttt{x2l}, containing the real-time observations of the functional response and covariates, divided between Phase I and Phase II. 

\begin{verbatim}
R> xl <- get_mfd_df_real_time(dt = dft,
+                             domain = c(0, 1),
+                             arg = "percent_miles_hat",
+                             id = "VN",
+                             variables = c(xvar, yvar),
+                             n_basis = 100)
R> y1l <- lapply(xl, function(x) x[obs1, yvar])
R> x1l <- lapply(xl, function(x) x[obs1, xvar])
R> y2l <- lapply(xl, function(x) x[c(obs1, obs2), yvar])
R> x2l <- lapply(xl, function(x) x[c(obs1, obs2), xvar])
\end{verbatim}

\renewcommand\refname{References}
\bibliography{refs.bib}